\documentclass[format=acmsmall,dvipsnames,table,nonacm]{acmart}

\setcopyright{none}
\renewcommand\footnotetextcopyrightpermission[1]{}
\usepackage{adjustbox}
\usepackage{supertabular, multicol}
\usepackage{booktabs}
\usepackage{amsmath}
\usepackage{afterpage}
\usepackage{pdflscape}
\usepackage[disable]{todonotes}
\usepackage{xspace}
\usepackage{xparse}
\usepackage[]{hyperref} %
\usepackage{listings}
\usepackage{color}
\usepackage[shortcuts]{extdash}
\usepackage[T1]{fontenc}
\usepackage{tabularx}
\usepackage{booktabs}
\usepackage[skip=0.5\baselineskip]{caption}
\usepackage{multirow}
\usepackage{rotating}
\usepackage{colortbl}
\usepackage[normalem]{ulem}
\useunder{\uline}{\ul}{}
\usepackage{adjustbox}
\usepackage[binary-units]{siunitx}
\usepackage{subcaption}

\usepackage{pifont}
\newcommand{\cmark}{{\color{Green}\ding{51}}}
\newcommand{\xmark}{{\color{Red}\ding{55}}}
\newcommand{\hcircle}{{\color{Orange}\ding{119}}}

\usepackage[inline]{enumitem}

\graphicspath{{./fig/}}
\DeclareGraphicsExtensions{.pdf}

\definecolor{mygreen}{rgb}{0,0.6,0}
\definecolor{mygray}{rgb}{0.5,0.5,0.5}
\definecolor{mymauve}{rgb}{0.58,0,0.82}
\colorlet{rowgray}{gray!15}
\colorlet{hlgreen}{Green!25}

\usepackage{array}
\newcolumntype{C}[1]{>{\centering\arraybackslash}m{#1}}

\def\greencell{{\cellcolor{hlgreen}}}

\DeclareSIUnit{\cpuhour}{CPU\textrm{-}hours}
\DeclareSIUnit{\loc}{LOC}

\ExplSyntaxOn
  \cs_new_eq:NN \calc \fp_eval:n
\ExplSyntaxOff

\def\sys{Magma\xspace}
\def\sysurl{\url{https://hexhive.epfl.ch/magma/}\xspace}
\def\lavam{LAVA\=/M\xspace}
\def\unifuzz{\textsc{UniFuzz}\xspace}
\def\libpngbugn{7}
\def\libtiffbugn{14}
\def\libxmlbugn{18}
\def\popplerbugn{22}
\def\opensslbugn{21}
\def\sqlitebugn{20}
\def\phpbugn{16}
\def\systargetcount{seven\xspace} %
\def\sysdrivercount{25\xspace}
\def\sysbugcount{\calc{\libpngbugn+\libtiffbugn+\libxmlbugn+\popplerbugn+\opensslbugn+\sqlitebugn+\phpbugn}\xspace} %
\def\syspovcount{54\xspace} %
\def\syscwecount{11\xspace}
\def\sysevalcount{seven\xspace} %
\def\sysevalruns{ten\xspace} %
\def\sysreachbugcount{74\xspace}
\def\systriggerbugcount{43\xspace}
\def\sysevalduration{\SI{24}{\hour}\xspace} %
\def\sysevallongduration{\SI{7}{\day}\xspace} %
\def\systotalcpuhours{\SI{200000}{\cpuhour}\xspace}
\def\colcombined{\textbf{Mean}\xspace}

\def\mopt{\textsc{MOpt}\xspace}
\def\fairfuzz{\textsc{FairFuzz}\xspace}
\def\symcc{\textsc{SymCC}\xspace}

\newlist{inlineroman}{enumerate*}{1}
\setlist[inlineroman]{label=(\roman*)}

\newlist{inlinealph}{enumerate*}{1}
\setlist[inlinealph]{label=(\alph*)}

\begin{document}

\title{\sys: A Ground-Truth Fuzzing Benchmark}

\author{{Ahmad Hazimeh}}
\affiliation{\institution{EPFL}\country{Switzerland}}
\email{ahmad.hazimeh@epfl.ch}
\author{{Adrian Herrera}}
\affiliation{\institution{ANU \& DST}\country{Australia}}
\email{adrian.herrera@anu.edu.au}
\author{{Mathias Payer}}
\affiliation{\institution{EPFL}\country{Switzerland}}
\email{mathias.payer@nebelwelt.net}

\begin{abstract}
High scalability and low running costs have made fuzz testing the de facto
standard for discovering software bugs. Fuzzing techniques are constantly being
improved in a race to build the ultimate bug-finding tool. However, while
fuzzing excels at finding bugs in the wild, evaluating and comparing fuzzer
performance is challenging due to the lack of metrics and benchmarks. For
example, crash count---perhaps the most commonly-used performance metric---is
inaccurate due to imperfections in deduplication techniques. Additionally, the
lack of a unified set of targets results in ad~hoc evaluations that hinder fair
comparison.

We tackle these problems by developing \emph{\sys}, a ground-truth fuzzing
benchmark that enables uniform fuzzer evaluation and comparison.
By introducing \emph{real} bugs into \emph{real} software, \sys allows for the
realistic evaluation of fuzzers against a broad set of targets. By instrumenting
these bugs, \sys also enables the collection of bug-centric performance
metrics independent of the fuzzer. \sys is an open benchmark consisting of
\systargetcount targets that perform a variety of input manipulations and
complex computations, presenting a challenge to state-of-the-art fuzzers.

We evaluate \sysevalcount widely-used mutation-based fuzzers (AFL, AFLFast,
AFL++, \fairfuzz, \mopt-AFL, honggfuzz, and \symcc-AFL) against \sys
over~\systotalcpuhours. Based on the number of bugs reached, triggered, and
detected, we draw conclusions about the fuzzers' exploration and detection
capabilities. This provides insight into fuzzer performance evaluation,
highlighting the importance of ground truth in performing more accurate and
meaningful evaluations.
\end{abstract}

\begin{CCSXML}
<ccs2012>
<concept>
<concept_id>10002944.10011123.10011124</concept_id>
<concept_desc>General and reference~Metrics</concept_desc>
<concept_significance>500</concept_significance>
</concept>
<concept>
<concept_id>10002944.10011123.10011130</concept_id>
<concept_desc>General and reference~Evaluation</concept_desc>
<concept_significance>500</concept_significance>
</concept>
<concept>
<concept_id>10011007.10011074.10011099.10011102</concept_id>
<concept_desc>Software and its engineering~Software defect analysis</concept_desc>
<concept_significance>500</concept_significance>
</concept>
<concept>
<concept_id>10002978.10003022</concept_id>
<concept_desc>Security and privacy~Software and application security</concept_desc>
<concept_significance>300</concept_significance>
</concept>
</ccs2012>
\end{CCSXML}

\ccsdesc[500]{General and reference~Metrics}
\ccsdesc[500]{General and reference~Evaluation}
\ccsdesc[500]{Software and its engineering~Software defect analysis}
\ccsdesc[300]{Security and privacy~Software and application security}

\keywords{fuzzing; benchmark; software security; performance evaluation}

\thanks{This project has received funding from the European Research Council
(ERC) under the European Union's Horizon 2020 research and innovation program
(grant agreement No. 850868)}

\maketitle
\thispagestyle{empty}
\renewcommand{\shortauthors}{A. Hazimeh, et al.}

\section{Introduction}
\label{sec:introduction}

Fuzz testing (``fuzzing'') is a widely-used dynamic bug discovery technique. A
fuzzer procedurally generates inputs and subjects the target program (the
``target'') to these inputs with the aim of triggering a fault (i.e.,
discovering a bug).
Fuzzing is an inherently sound but incomplete bug-finding process (given finite
resources).
State-of-the-art fuzzers rely on \emph{crashes} to
mark faulty program behavior.
The existence of a crash is generally symptomatic
of a bug (soundness), but the lack of a crash does not necessarily mean that the
program is bug-free (incompleteness).
Fuzzing is wildly successful in finding bugs in
open-source~\cite{ossfuzz} and commercial off-the-shelf~\cite{adobefuzz,
msftfuzz, gfuzzforsec} software.

The success of fuzzing has resulted in an explosion of new techniques claiming
to improve bug-finding performance~\cite{fuzzingart}. In order to highlight
improvements, these techniques are typically evaluated across a range of
metrics, including:
\begin{inlineroman}
\item crash counts;
\item ground-truth bug counts; and/or
\item code-coverage profiles.
\end{inlineroman}
While these metrics provide some insight into a fuzzer's performance, we argue
that they are insufficient for use in fuzzer comparisons. Furthermore, the set
of targets that these metrics are evaluated on can vary wildly across papers,
making cross-fuzzer comparisons impossible. Each of these metrics has particular
deficiencies.

\paragraph*{Crash counts}
The simplest fuzzer evaluation method is to count the number of crashes
triggered by a fuzzer, and compare this crash count with that achieved by
another fuzzer (on the same target). Unfortunately, crash counts often inflate
the number of actual bugs in the target~\cite{fuzzeval}. Moreover, deduplication
techniques (e.g., coverage profiles, stack hashes) fail to accurately identify
the root cause of these crashes~\cite{fuzzeval, aurora}.

\paragraph*{Bug counts}
Identifying a crash's \emph{root cause} is preferable to simply reporting
raw crashes, as it avoids the inflation problem inherent in crash counts.
Unfortunately, obtaining an accurate \emph{ground-truth} bug count typically
requires extensive manual triage, which in turn requires someone with extensive
domain expertise and experience~\cite{industrialhacking}.

\paragraph*{Code-coverage profiles}
Code-coverage profiles are another performance metric commonly used to evaluate
and compare fuzzing techniques. Intuitively, covering more code correlates with
finding more bugs. However, previous work~\cite{fuzzeval} has shown that there
is a weak correlation between coverage-deduplicated crashes and ground-truth
bugs, implying that higher coverage does not necessarily indicate better fuzzer
effectiveness.

\medskip

The deficiencies of existing performance metrics calls for a rethinking of fuzzer
evaluation practices. In particular, the performance metrics used in these
evaluations must accurately measure a fuzzer's ability to achieve its main
objective: \emph{finding bugs}. Similarly, the targets that are used to assess
how well a fuzzer meets this objective must be realistic and exercise diverse
behavior. This allows a practitioner to have confidence that a given fuzzing
technique will yield improvements when deployed in real-world environments.

To satisfy these criteria, we present \emph{\sys}, a ground-truth fuzzer
benchmark based on real programs with real bugs. \sys consists of
\systargetcount widely-used open-source libraries and applications, totalling
\SI{2}{\mega\loc}. For each \sys workload, we manually analyze security-relevant
bug reports and patches, reinserting defective code back into these
\systargetcount programs (in total,~\sysbugcount bugs were analyzed and
reinserted). Additionally, each reinserted bug is accompanied by a light-weight
\emph{oracle} that detects and reports if the bug is \emph{reached} or
\emph{triggered}. This distinction between reaching and triggering a bug---in
addition to a fuzzer's ability to \emph{detect} a triggered bug---presents a new
opportunity to evaluate a fuzzer across multiple dimensions (again, focusing on
ground-truth bugs).

The remainder of this paper presents the motivation behind \sys, the methodology
behind \sys's design and choice of performance metrics, implementation details,
and a set of preliminary results that demonstrate \sys's utility.
We make the following contributions:
\begin{itemize}
\item A set of bug-centric performance metrics for a
fuzzer benchmark that allow for a fair and accurate evaluation and comparison of
fuzzers.

\item A quantitative comparison of existing fuzzer benchmarks.

\item The design and implementation of \sys, a ground-truth fuzzing benchmark
based on real programs with real bugs.

\item An evaluation of \sys against \sysevalcount widely-used
fuzzers.
\end{itemize}

\section{Background and Motivation}
\label{sec:background}

This section introduces fuzzing as a software testing technique, and how new
fuzzing techniques are currently evaluated and compared against existing ones.
This aims to motivate the need for new fuzzer evaluation practices.

\subsection{Fuzz testing (fuzzing)}
\label{sec:fuzzing-background}

A fuzzer is a dynamic testing tool that discovers software flaws by running a
target program (the ``target'') with a large number of automatically-generated
inputs. Importantly, these inputs are generated with the intention of triggering
a crash in the target. This input generation process is dependent on the
fuzzer's knowledge of the target's \emph{input format} and \emph{program
structure}.
For example, \emph{grammar-based} fuzzers (e.g., Superion~\cite{superion},
Peachfuzz~\cite{peachfuzz}, and QuickFuzz~\cite{quickfuzz}) leverage the
target's input format (which must be specified \textit{a priori}) to
intelligently craft inputs (e.g., based on data width and type, and on the
relationships between different input fields). In contrast, \emph{mutational}
fuzzers (e.g., AFL~\cite{afl}, Angora~\cite{angora}, and MemFuzz~\cite{memfuzz})
require no \textit{a priori} knowledge of the input format. Instead, mutational
fuzzers leverage preprogrammed mutation operations to iteratively modify the
input.

Fuzzers are classified by their knowledge of the target's program
structure. For example, \emph{whitebox} fuzzers~\cite{buzzfuzz, gode2008, mowf}
leverage program analysis to infer knowledge about the program structure. In
comparison, \emph{blackbox} fuzzers~\cite{radamsa, fuzzsim} blindly generate
inputs in the hope of discovering a crash. Finally, \emph{greybox}
fuzzers~\cite{afl, angora, libfuzzer} leverage program instrumentation (instead
of program analysis) to collect runtime information. Program-structure knowledge
guides input generation in a manner more likely to trigger a crash.

Importantly, fuzzing is a \emph{highly stochastic} bug-finding process. This
randomness is independent of whether the fuzzer synthesizes inputs from a
grammar (grammar-based fuzzing), transforms an existing set of inputs to arrive
at new inputs (mutational fuzzing), has no knowledge of that target's internals
(blackbox fuzzing), or uses sophisticated program analyses to understand the 
target (whitebox fuzzing). The stochastic nature of fuzzing makes evaluating and
comparing fuzzers difficult. This problem is exacerbated by existing fuzzer
evaluation metrics and benchmarks.

\subsection{The Current State of Fuzzer Evaluation}
\label{sec:current-fuzzer-eval}

The rapid emergence of new and improved fuzzing techniques~\cite{fuzzingart}
means that fuzzers are constantly compared against one another, in order to
empirically demonstrate that the latest fuzzer supersedes previous
state-of-the-art fuzzers. To enable fair and accurate fuzzer evaluation, it is
critical that fuzzing campaigns are conducted on a suitable benchmark that uses
an appropriate set of metrics.
Unfortunately, fuzzer evaluations have so far been ad hoc and haphazard. For
example, Klees et al.'s study of~32 fuzzing papers found that \emph{none} of
the surveyed papers provided sufficient detail to support their claims of fuzzer
improvement~\cite{fuzzeval}. Notably, their study highlights a set of criteria
that should be adopted across all fuzzer evaluations. These criteria include:
\begin{description}
\item[Performance metrics:] How the fuzzers are evaluated and compared. This is
typically one of the approaches previously discussed (crash count, bug count, or
coverage profiling).

\item[Targets:] The software being fuzzed. This software should be both diverse
and realistic so that a practitioner has confidence that the fuzzer will perform
similarly in real-world environments.

\item[Seed selection:] The initial set of inputs that bootstrap the fuzzing
process. This initial set of inputs should be consistent across repeated trials
and the fuzzers under evaluation.

\item[Trial duration (timeout):] The length of a single fuzzing trial should
also be consistent across repeated trials and the fuzzers under evaluation. We
use the term \emph{trial} to refer to an instance of the fuzzing process on a
target program, while a \emph{fuzzing campaign} is a set of~$N$ repeated trials
on the same target.

\item[Number of trials:] The highly-stochastic nature of fuzzing necessitates a
large number of repeated trials, allowing for a statistically sound comparison
of results.
\end{description}

Klees et al.'s study demonstrates the need for a \emph{ground-truth fuzzing
benchmark}. Such a benchmark must use suitable performance metrics and present
a unified set of targets.

\subsubsection{Existing Fuzzer Benchmarks}
\label{sec:existing-benchmarks}

Fuzzers are typically evaluated on a set of targets sourced from one of the
following benchmarks. These benchmarks are summarized in
\autoref{tab:existing-benchmarks}.

\begin{table}[b]
\footnotesize
\centering

\caption{Summary of existing fuzzer benchmarks and our benchmark, \sys. We
characterize benchmarks across two dimensions: the targets that make up the
benchmark workloads and the bugs that exist across these workloads. For both
dimensions we count the number of workloads/bugs (\#) and classify them as
\textbf{R}eal or \textbf{S}ynthetic. Bug density is the mean number of bugs per
workload. Finally, ground truth may be available (\cmark), available but not
easily accessible (\hcircle), or unavailable (\xmark).}
\label{tab:existing-benchmarks}

\rowcolors{3}{rowgray}{white}
\begin{tabular}{l|rc|rc|rc}
  \toprule
  \multirow{2}{*}{Benchmark} &
  \multicolumn{2}{c|}{Workloads} &
  \multicolumn{2}{c|}{Bugs}  &
  \multirow{2}{*}{Bug Density} &
  \multirow{2}{*}{Ground truth} \\
                                    & \#        & Real/Synthetic    & \#         & Real/Synthetic &              & \\
  \midrule
  BugBench~\cite{bugbench}          & 17        & R                 & 19         & R              & \num{1.12}   & \hcircle \\
  CGC~\cite{cgc}                    & 131       & S                 & 590        & S              & \num{4.50}   & \hcircle \\
  Google FTS~\cite{gfuzzsuite}      & 24        & R                 & 47         & R              & \num{1.96}   & \hcircle \\
  Google FuzzBench~\cite{fuzzbench} & 21        & R                 & \textminus & \textminus     & \textminus   & \textminus \\
  \lavam~\cite{lavam}               & 4         & R                 & 2265       & S              & \num{566.25} & \cmark \\
  \unifuzz~\cite{unifuzz}           & 20        & R                 & ?          & R              & ?            & \xmark \\
  Open-source software              & \textminus& R                 & ?          & R              & ?            & \xmark \\
  \midrule
  \sys                              & 7         & R                 & \sysbugcount & R           & \num{16.86}  & \cmark \\
  \bottomrule
\end{tabular}
\end{table}

The \lavam~\cite{lavam} test suite (built on top of \texttt{coreutils-8.24})
aims to evaluate the effectiveness of a fuzzer's exploration capability by
injecting bugs in different execution paths. However, the LAVA bug injection
technique only injects a single, simple bug type: an out-of-bounds memory access
triggered by a ``magic value'' comparison. This bug type does not accurately
represent the statefulness and complexity of bugs encountered in real-world
software. We quantify these observations in \autoref{sec:lavam-evaluation}.

In contrast, the Cyber Grand Challenge (CGC)~\cite{cgc} sample set provides a
wider variety of bugs that are suitable for testing a fuzzer's fault detection
capabilities. Unfortunately, the relatively small size and simplicity of the
CGC's synthetic workloads does not enable thorough evaluation of the fuzzer's
ability to explore complex programs.

BugBench~\cite{bugbench} and the Google Fuzzer Test Suite
(FTS)~\cite{gfuzzsuite} both contain real programs with real bugs. However, each
target only contains one or two bugs (on average). This sparsity of bugs,
combined with the lack of automatic methods for triaging crashes, hinders
adoption and makes both benchmarks unsuitable for fuzzer evaluation.
In contrast, Google FuzzBench~\cite{fuzzbench}---the successor to the Google
FTS---is a fuzzer evaluation platform that relies solely on coverage profiles as
a performance metric. As previously discussed, this metric has limited utility
when evaluating fuzzers on their bug-finding capability.
\unifuzz~\cite{unifuzz}---which was developed concurrently but independently
from \sys---is similarly built on real programs containing real bugs. However,
it lacks ground-truth knowledge and it is unclear how many bugs each target
contains. Not knowing how many bugs exist in a benchmark makes fuzzer
comparisons challenging.

Finally, popular open-source software (OSS) is often used to evaluate
fuzzers~\cite{fuzzeval, fairfuzz, aflfast, moptafl, tfuzz, skyfire}. Although
real-world software is used, the lack of ground-truth knowledge about the
triggered crashes makes it difficult to provide an accurate, verifiable,
quantitative evaluation. First, it is often unclear which software version is
used, making fair cross-paper comparisons impossible.
Second, multiple software versions introduce \emph{version divergence}, a subtle
evaluation flaw shared by both crash and bug count metrics. After running for an
extended period, a fuzzer's ability to discover new bugs diminishes over
time~\cite{expcostbugs}. If a second fuzzer later fuzzes a new version of the
same program---with the bugs found by the first fuzzer appropriately
patched---then the first fuzzer will find fewer bugs in this newer version.
Version divergence is also inherent in \unifuzz, which builds on top of older
versions of OSS.

\subsubsection{Crashes as a Performance Metric}
\label{sec:crash-count}

Most, if not all, state-of-the-art fuzzers implement fault detection as a
\emph{crash listener}. A program crash can be caused by an \emph{architectural
violation} (e.g., division-by-zero, unmapped/unprivileged page access) or by a
\emph{sanitizer} (a dynamic bug-finding tool that generates a crash when a
security policy violation---e.g., object out-of-bounds, type safety
violation---occurs~\cite{sanitizers}).

The simplicity of crash detection has led to the widespread use of \emph{crash
count} as a performance metric for comparing fuzzers. However, crash counts have
been shown to yield inflated results, even when combined with deduplication
methods (e.g., coverage profiles and stack hashes)~\cite{fuzzeval, aurora}.
Instead, the number of bugs found by each fuzzer should be compared: if
fuzzer~$A$ finds more bugs than fuzzer~$B$, then~$A$ is superior to~$B$.
Unfortunately, there is no single formal definition for a bug. Defining a bug in
its proper context is best achieved by formally modeling program behavior.
However, deriving formal program models is a difficult and time-consuming task.
As such, bug detection techniques tend to create a blacklist of faulty behavior,
mislabeling or overlooking some bug classes in the process. This often leads to
incomplete detection of bugs and root-cause misidentification, resulting in a
duplication of crashes and an inflated set of results.

\section{Desired Benchmark Properties}
\label{sec:benchmarks}

Benchmarks are important drivers for computer science research and product
development~\cite{dacapo}.  Several factors must be taken into account when
designing a benchmark, including: relevance; reproducibility; fairness;
verifiability; and usability~\cite{howtobuildabenchmark, bvat}. While building
benchmarks around these properties is well studied~\cite{howtobuildabenchmark,
dacapo, renaissance, speccpu2000, bugbench, bvat, specbms, buginjector,
bug-synthesis}, the highly-stochastic nature of fuzzing introduces new
challenges for benchmark designers.

For example, \emph{reproducibility} is a key benchmark property that ensures a
benchmark produces ``\textit{the same results consistently for a particular test
environment}''~\cite{howtobuildabenchmark}. However, individual fuzzing trials
vary wildly in performance, requiring a large number of repeated trials for a
particular test environment~\cite{fuzzeval}. While performance variance exists
in most benchmarks (e.g., the SPEC CPU benchmark~\cite{specbms} uses the median
of three repeated trials to account for small variations across environments),
this variance is more pronounced in fuzzing. Furthermore, a fuzzer may actively
modify the test environment (e.g., T-Fuzz~\cite{tfuzz} and
FuzzGen~\cite{fuzzgen} transform the target, while Skyfire~\cite{skyfire}
generates new seed inputs for the target). This is very different to traditional
performance benchmarks (e.g., SPEC CPU~\cite{specbms}, DaCapo~\cite{dacapo}),
where the workloads and their inputs remain fixed across all systems-under-test.
This leads us to define the following set of properties that we argue
\emph{must} exist in a fuzzing benchmark:
\begin{description}
\item[Diversity (P1):] The benchmark contains a wide variety of bugs and
programs that resemble real software testing scenarios.

\item[Verifiability (P2):] The benchmark yields verifiable metrics that
accurately describe performance.

\item[Usability (P3):] The benchmark is accessible and has no significant
barriers for adoption.
\end{description}
These three properties are explored in the remainder of this section, while
\autoref{sec:approach} describes how \sys satisfies these criteria.

\subsection{Diversity (P1)}
\label{sec:diversity}

Fuzzers are actively used to find bugs in a variety of \emph{real}
programs~\cite{ossfuzz, adobefuzz, msftfuzz, gfuzzforsec}. Therefore, a fuzzing
benchmark must evaluate fuzzers against programs and bugs that resemble those
encountered in the ``real world''. To this end, a benchmark must include a
\emph{diverse} set of bugs \emph{and} programs.

Bugs should be diverse with respect to:
\begin{description}
\item[Class:] Common Weakness Enumeration (CWE)~\cite{cwe} bug classes include
memory-based errors, type errors, concurrency issues, and numeric errors.

\item[Distribution:] ``Depth'', fan-in (i.e, the number of paths which execute
the bug), and spread (i.e., the ratio of faulty-path counts to the total number
of paths).

\item[Complexity:] Number of input bytes involved in triggering a bug, the range
of input values which triggers the bug, and the transformations performed on the
input.
\end{description}

Similarly, targets (i.e, the benchmark workloads) should be diverse with respect
to:
\begin{description}
\item[Application domain:] File and media processing, network protocols,
document parsing, cryptography primitives, and data encoding.

\item[Operations performed:] Parsing, checksum calculation, indirection,
transformation, state management, and data validation.

\item[Input structure:] Binary, text, formats/grammars, and data size.
\end{description}

Satisfying the diversity property requires bugs that resemble those encountered in
real-world environments. Both \lavam and Google FuzzBench fail this requirement:
the former contains only a single bug class (an out-of-bounds memory access),
while FuzzBench does not consider bugs as an evaluation metric.
BugBench primarily focuses on memory corruption vulnerabilities, but also
contains uninitialized read, memory leak, data race, atomicity, and semantic
bugs (totalling nine bug classes).
Conversely, Google FTS and FuzzBench satisfy the target diversity requirement:
both contain workloads from a wide variety of application domains (e.g.,
cryptography, image parsing, text processing, and compilers).

Ultimately, real programs are the only source of real bugs. Therefore, a
benchmark designed to evaluate fuzzers must include \emph{real programs with a
variety of real bugs}, thus ensuring diversity and avoiding bias (e.g., towards
a specific bug class). Whereas discovering and reporting real bugs is desirable
(i.e, when OSS is used), performance metrics based on an unknown set of bugs
(with an unknown distribution) make it impossible to compare fuzzers. Instead,
fuzzers should be evaluated on workloads containing known bugs for which ground
truth is available and \emph{verifiable}.

\subsection{Verifiability (P2)}

Existing ground-truth fuzzing benchmarks lack a straightforward mechanism for
determining a crash's root cause. This makes it difficult to verify a fuzzer's
results. Crash count, a widely-used performance metric, suffers from high
variability, double-counting, and inconsistent results across multiple trials
(see \autoref{sec:crash-count}). Automated techniques for deduplicating crashes
are not reliable, and hence should not be used to verify the bugs discovered by
a fuzzer. Ultimately, a fuzzing benchmark should provide a set of known bugs for
which ground truth can be used to verify a fuzzer's findings.

While the CGC sample set provides crashing inputs---also known as a \emph{proof
of vulnerability} (PoV)---for all known bugs, it does not provide a mechanism
for determining the root cause of a fuzzer-generated crash.
Similarly, the Google FTS provides PoVs (for \SI{87}{\percent} of bugs) and a
script for triaging and deduplicating crashes. This script parses the crash
report or looks for a specific line of code at which to terminate program
execution. However, this approach is limited and does not allow for the
detection of complex bugs (e.g., where simply executing a line of code is not
sufficient to trigger the bug).

In contrast to the CGC and Google FTS benchmarks, for which ground truth is
available but not easily accessible, \lavam clearly reports the bug triggered by
a crashing input. However, \lavam does not provide a runtime interface for
accessing this information. Unless a fuzzer is specialized to collect \lavam
metrics, it cannot monitor progress in real-time. Thus, a post-processing step
is required to collect metrics.
Finally, Google FuzzBench relies solely on coverage profiles (rather than
fault-based metrics) to evaluate and compare fuzzers. FuzzBench dismisses the
need for ground truth, which we believe sacrifices the significance of the
results: more coverage does not necessarily imply higher bug-finding
effectiveness.

Ground-truth bug knowledge allows for a fuzzer's findings to be verified,
enabling accurate performance evaluation and allowing meaningful comparisons
between fuzzers. To this end, a fuzzing benchmark must provide \emph{easy access
to ground-truth metrics} describing the bugs a fuzzer can reach, trigger, and
detect.

\subsection{Usability (P3)}

Fuzzers have evolved from simple blackbox random-input generation to complex
control- and data-flow analysis tools. Each fuzzer may introduce its own
instrumentation into a target (e.g., AFL~\cite{afl}), run the target in a
specific execution engine (e.g., QSYM~\cite{qsym}, Driller~\cite{driller}), or
provide inputs through a specific channel (e.g., libFuzzer~\cite{libfuzzer}).
Fuzzers come in a variety of forms (described in
\autoref{sec:fuzzing-background}), so a fuzzing benchmark must not exclude a
particular type of fuzzer. Additionally, using a benchmark must be manageable
and straightforward: it should not require constant user intervention, and
benchmarking should finish within a reasonable time frame. The inherent
randomness of fuzzing complicates this, as multiple trials are required to
achieve statistically-meaningful results.

Some existing benchmark workloads (e.g., those from CGC and Google FTS) contain
multiple bugs, so it is not sufficient to only run the fuzzer until the first
crash is encountered. However, the lack of easily-accessible ground truth makes
it difficult to determine if/when all bugs are triggered. Moreover, inaccurate
deduplication techniques mean that the user cannot simply equate the number of
crashes with the number of bugs. Thus, additional time must be spent triaging
crashes to obtain ground-truth bug counts, further complicating the benchmarking
process.

In summary, a benchmark should be \emph{usable} by fuzzer developers, without
introducing insurmountable or impractical barriers to adoption. To satisfy this
property, a benchmark must thus provide a \emph{small set of targets with a
large number of discoverable bugs}, and it must provide a \emph{usable framework
that measures and reports fuzzer progress and performance}.

\section{\sys: Approach}
\label{sec:approach}

We present \sys, a ground-truth fuzzing benchmark that satisfies the
previously-discussed benchmark properties. \sys is a collection of
\systargetcount targets with widespread use in real-world environments. These
initial targets have been carefully selected for their \emph{diversity} and the
variety of security-critical bugs that have been reported throughout their
lifetimes (satisfying \textbf{P1}).

Importantly, \sys's \systargetcount workloads contain~\sysbugcount bugs for
which ground truth is \emph{easily accessible} and \emph{verifiable} (satisfying
\textbf{P2}). These bugs are sourced from older versions of the \systargetcount
workloads, and then \emph{forward-ported} to the latest version contained within
\sys. Finally, \sys imposes minimal requirements on the user, allowing fuzzer
developers to seamlessly integrate the benchmark into their development cycle
(satisfying \textbf{P3}).

\label{sec:overview}

For each workload, we manually inspect bug and vulnerability reports to find
bugs that are suitable for inclusion in \sys (e.g., ensuring that the bug
affects the core codebase). For these bugs, we reintroduce (``inject'') each bug
into the latest version of the code through a process we call
\emph{forward-porting} (see \autoref{sec:bug-selection}). In addition to the
bug, we also insert minimal source-code instrumentation---a \emph{canary}---to
collect data about a fuzzer's ability to reach and trigger the bug (see
\autoref{sec:performance-metrics}). A bug is \emph{reached} when the faulty line
of code is executed, and \emph{triggered} when the fault condition is satisfied.
Finally, \sys provides a \emph{runtime monitor} that runs in parallel with the
fuzzer to collect real-time statistics. These statistics are used to evaluate
the fuzzer (see \autoref{sec:runtime-monitor}).

Fuzzer evaluation is based on the number of bugs \emph{reached},
\emph{triggered}, and \emph{detected}. The \sys instrumentation only yields
usable information when the fuzzer exercises the instrumented code, allowing us
to determine whether a bug is \emph{reached}. The fuzzer-generated input
\emph{triggers} a bug when the input's dataflow satisfies the bug's trigger
condition(s). Once triggered, the fuzzer should flag the bug as a fault or
crash, enabling us to assess the fuzzer's bug \emph{detection} capability. These
metrics are described further in \autoref{sec:performance-metrics}.

Finally, \sys provides a \emph{fatal canaries} mode. In fatal canaries mode, the
program is terminated if a canary's condition is satisfied (similar to \lavam).
The fuzzer then saves this crashing input for post-processing. Fatal canaries
are a form of \emph{ideal sanitization}, in which triggering a bug immediately
results in a crash, regardless of the nature of the bug.
Fatal canaries allow developers to evaluate their fuzzers under ideal
sanitization assumptions without incurring additional sanitization overhead.
This mode increases the number of executions during an evaluation, reducing the
cost of evaluating a fuzzer but sacrificing the ability to evaluate a fuzzer's
detection capabilities.

\subsection{Target Selection}
\label{sec:target-selection}

\sys contains \systargetcount targets, which we summarize in
\autoref{tab:target-summary}. In addition to these \systargetcount
\emph{targets} (i.e., the codebases into which bugs are injected), \sys also
includes~\sysdrivercount \emph{drivers} (i.e., executable programs that provide
a command-line interface to the target) that exercise different functionality
within the target. Inspired by Google OSS-Fuzz~\cite{ossfuzz}, these drivers are
sourced from the original target codebases (as drivers are best developed by
domain experts).

\begin{table}[b]
\centering
\caption{The targets, driver programs, bug counts, and evaluated features
incorporated into \sys. The versions used are the latest at the time of
writing.}
\label{tab:target-summary}

\begin{adjustbox}{width=\linewidth}
  \rowcolors{2}{white}{rowgray}
  \begin{tabular}{l>{\raggedright}m{40mm}m{16mm}lrC{12mm}C{16mm}ccC{12mm}}
  \toprule
  Target              & Drivers                                                                              & Version              & File type           & Bugs         & Magic values & Recursive parsing & Compression & Checksums & Global state \\
  \midrule
  \textit{libpng}     & \texttt{read\_fuzzer}, \texttt{readpng}                                              & 1.6.38               & PNG                 & \libpngbugn  & \cmark       & \xmark            & \cmark      & \cmark    & \xmark       \\
  \textit{libtiff}    & \texttt{read\_rgba\_fuzzer}, \texttt{tiffcp}                                         & 4.1.0                & TIFF                & \libtiffbugn & \cmark       & \xmark            & \cmark      & \xmark    & \xmark       \\
  \textit{libxml2}    & \texttt{read\_memory\_fuzzer}, \texttt{xml\_reader\_for\_file\_fuzzer},
                        \texttt{xmllint}                                                                     & 2.9.10               & XML                 & \libxmlbugn  & \cmark       & \cmark            & \xmark      & \xmark    & \xmark       \\
  \textit{poppler}    & \texttt{pdf\_fuzzer}, \texttt{pdfimages}, \texttt{pdftoppm}                          & 0.88.0               & PDF                 & \popplerbugn & \cmark       & \cmark            & \cmark      & \cmark    & \xmark       \\
  \textit{openssl}    & \texttt{asn1}, \texttt{asn1parse}, \texttt{bignum}, \texttt{bndiv},
                        \texttt{client}, \texttt{cms}, \texttt{conf}, \texttt{crl},
                        \texttt{ct}, \texttt{server}, \texttt{x509}                                          & 3.0.0                & \emph{Binary blobs} & \opensslbugn & \cmark       & \xmark            & \cmark      & \cmark    & \cmark       \\
  \textit{sqlite3}    & \texttt{sqlite3\_fuzz}                                                               & 3.32.0               & SQL queries         & \sqlitebugn  & \cmark       & \cmark            & \xmark      & \xmark    & \cmark       \\
  \textit{php}        & \texttt{exif}, \texttt{json}, \texttt{parser}, \texttt{unserialize}                  & 8.0.0{\textminus}dev & \emph{Various}      & \phpbugn     & \cmark       & \cmark            & \xmark      & \xmark    & \xmark       \\ \bottomrule
  \end{tabular}
\end{adjustbox}

\end{table}

\sys's \systargetcount targets were selected for their diversity in
functionality (summarized qualitatively in \autoref{tab:target-summary}).
Inspired by benchmarks in other
fields~\cite{dacapo,renaissance,benchmarksimilarity,programsimilarity}, we apply
\emph{Principal Component Analysis} (PCA) to quantify this diversity. PCA is a
statistical analysis technique that transforms an $N$-dimensional space into a
lower-dimensional space while preserving variance as much as
possible~\cite{pca}. Reducing high-dimensional data into a set of
\emph{principal components} allows for the application of visualization and/or
clustering techniques to compare and discriminate benchmark workloads.

\begin{figure}
  \centering
  \begin{subfigure}{\linewidth}
    \centering
    \includegraphics[width=0.5\linewidth,keepaspectratio]{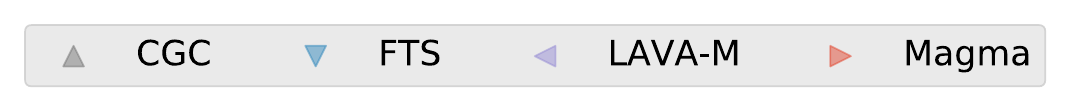}
  \end{subfigure}
  \begin{subfigure}{0.49\linewidth}
    \centering
    \includegraphics[width=\linewidth,keepaspectratio]{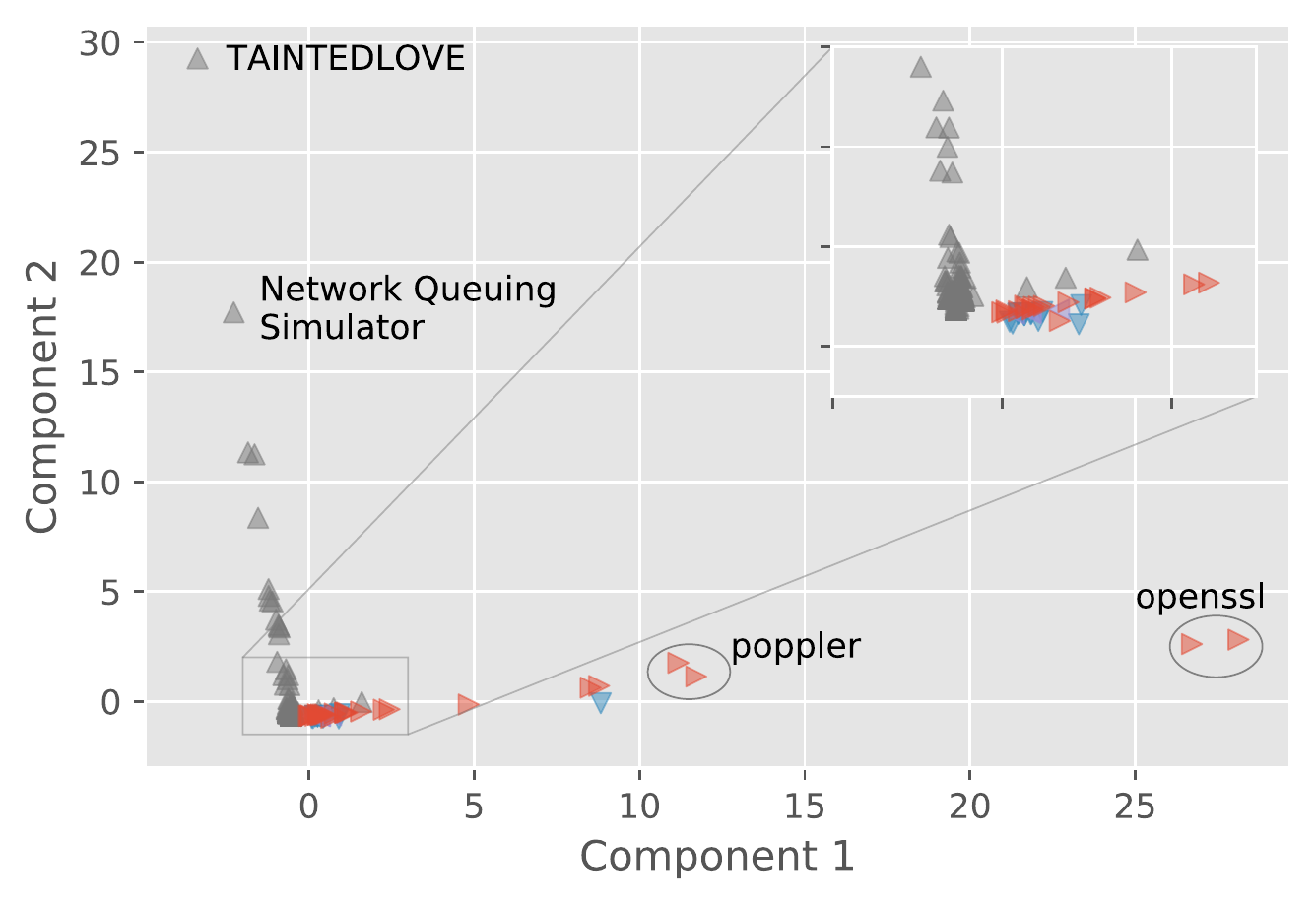}
  \end{subfigure}
  \begin{subfigure}{0.49\linewidth}
    \centering
    \includegraphics[width=\linewidth,keepaspectratio]{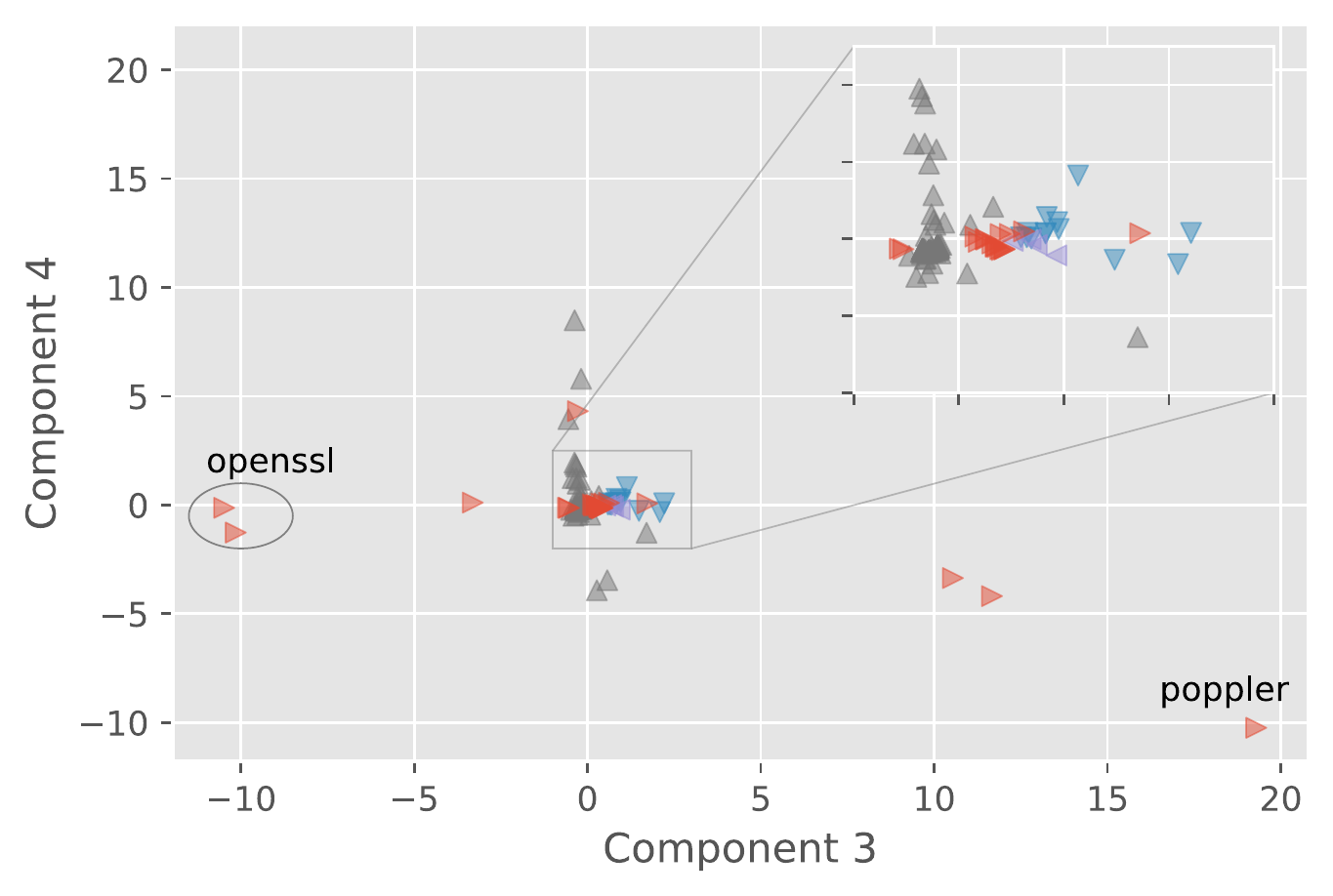}
  \end{subfigure}

  \caption{Scatter plots of benchmark scores over the first four principal
  components (which account for~$\sim$\SI{60}{\percent} of the variance in the
  benchmark workloads). Each point corresponds to a particular subject in a
  benchmark.}
  \label{fig:target-pca}
\end{figure}

We apply PCA as follows. First, we use an Intel Pin~\cite{intelpin} tool to
record instruction traces for $K=284$ \emph{subjects} (i.e., a library wrapped
with a particular driver program~\cite{ankou,libfuzzer}): four from \lavam,~14
from the FTS,~\sysdrivercount from \sys, and~241 from the CGC~\cite{cb-multios}.
Each trace is driven by seeds provided by the benchmark (exercising
functionality---and hence code---that would be explored by a fuzzer) and
contains instructions executed by both the subject and any linked libraries.
Second, instructions are categorized according to Intel XED, a disassembler
built into Pin. A XED instruction category is ``\textit{a higher level semantic
description of an instruction than its opcodes}''~\cite{intelpinapi}. XED
contains~$N=94$ instruction categories, spanning logical, floating point,
syscall, and SIMD operations (amongst others). We use these categories as an
approximation of the subject's functionality.
Third, we create a matrix~$X$, where~$x_{ij} \in X$ ($i \in [1, N]$
and~$j \in [1, K]$) is the mean number of instructions executed in a particular
category for a given subject (over all seeds supplied with that subject).
Finally, PCA is performed on a normalized version of~$X$. The first four
principal components, which in our case account for~\SI{60}{\percent} of the
variance between benchmarks, are plotted in a two-dimensional space in
\autoref{fig:target-pca}.

\autoref{fig:target-pca} shows that the four \lavam workloads are tightly
clustered over the first four principal components. This is unsurprising, given
that the \lavam workloads are all sourced from coreutils and hence share the same
codebase. In contrast, both the CGC and \sys provide a wide-variety of
workloads. For example, \textit{openssl}---which contains a large amount of
cryptographic and networking code---appears distinct from the main clusters in
\autoref{fig:target-pca}. The CGC's \emph{TAINTEDLOVE} workload is similarly
distinct, due to the relatively large number of floating point operations
performed.

\subsection{Bug Selection and Insertion}
\label{sec:bug-selection}

\begin{figure}
\centering
\includegraphics[width=0.6\linewidth,keepaspectratio]{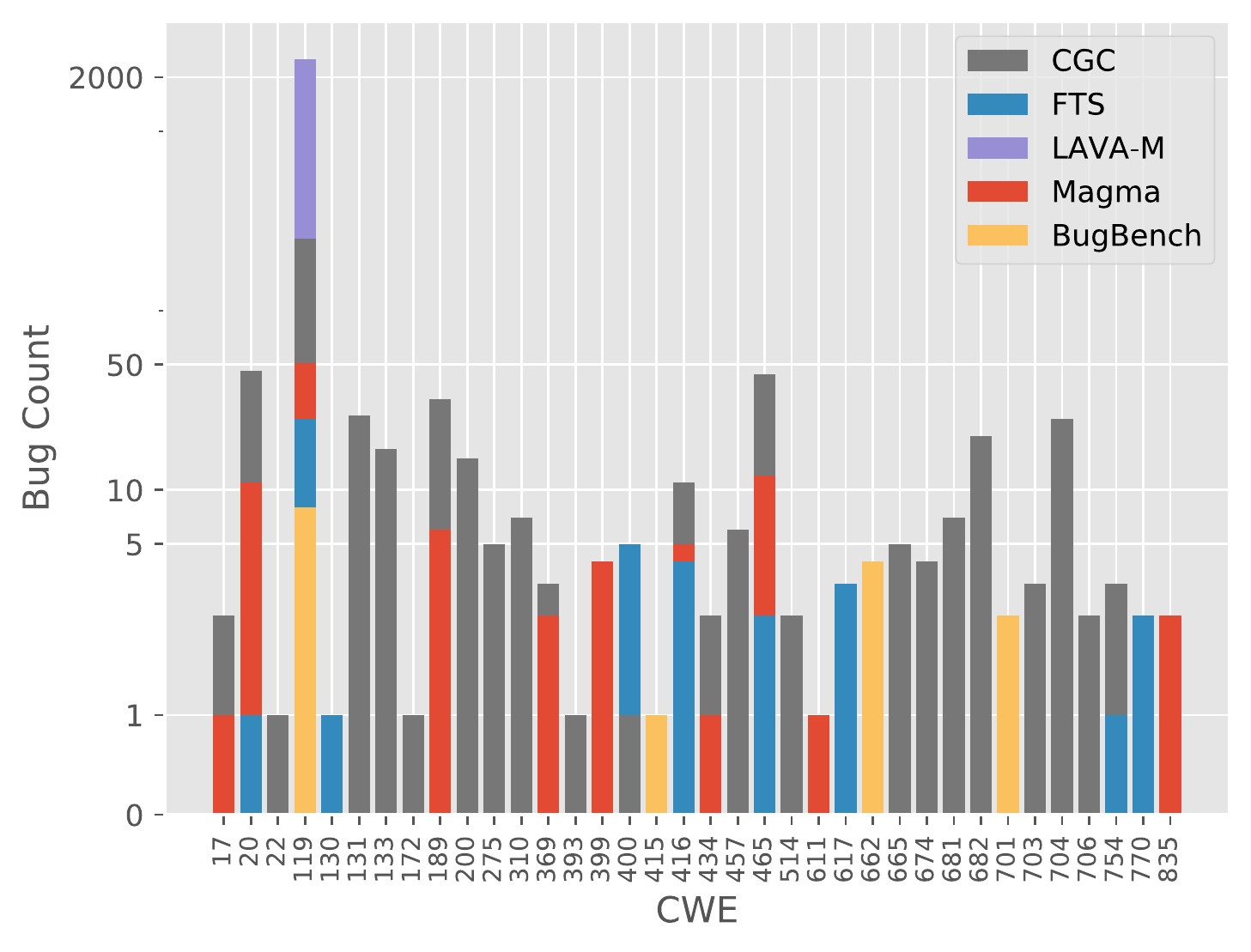}

\caption{Comparison of benchmark bug classes. The $y$-axis uses a log scale. A
complete list of \sys bugs is presented in \autoref{tab:all-bugs}.}
\label{fig:bug-summary}
\end{figure}

\sys contains~\sysbugcount bugs, spanning~\syscwecount CWEs (summarized
in \autoref{fig:bug-summary}; the complete list of bugs is given in
\autoref{tab:all-bugs}). Compared to existing benchmarks, \sys has both the
second-largest variety of bugs (by CWE) and second-largest ``bug density'' (the
ratio of the number of bugs to the number of targets) after the CGC and \lavam,
respectively. While the CGC has a wider variety of bugs, its workloads are not
indicative of real-world software (in terms of both size and complexity).
Similarly, while \lavam's bug density (\num{566.25} bugs per target) is an
order-of-magnitude larger than \sys's (\num{16.86} bugs per target), \lavam is
restricted to a single, synthetic bug type.

Importantly, \sys contains \emph{real} bugs sourced from bug reports and
\emph{forward-ported} to the most recent version of the target codebase. This is
in contrast to existing fuzzing benchmarks (e.g., BugBench, Google FTS) that
rely on old, unpatched versions of the target codebase. Unfortunately, using
older codebases limits the number of bugs available in each target (as evident
by the low bug densities in \autoref{tab:existing-benchmarks}). In comparison,
forward-porting---which is synonymous to \emph{back-porting} fixes from newer
codebases to older, buggy releases---does not suffer from this issue, making
\sys's targets \emph{easily extensible}.

Forward-porting begins with the identification---from the reported bug fix---of
the code changes that must be reverted to reintroduce the bug. Bug-fix commits
can contain multiple fixes to one or more bugs, so disambiguation is necessary
to prevent the introduction of unintended bugs. Alternatively, bug fixes may be
spread over multiple commits (e.g., if the original fix did not cover all edge
cases).
Following the identification of code changes, we identify what program state is
involved in evaluating the trigger condition. If necessary, we introduce
additional program variables to access that state. From this state, we determine
a boolean expression that serves as a light-weight oracle for identifying a triggered bug.
Finally, we identify a point in the program where we inject a canary before the
bug can manifest faulty behavior. This canary helps measure our fuzzer
performance metrics, discussed in the following section.

\subsection{Performance Metrics}
\label{sec:performance-metrics}

Fuzzer evaluation has traditionally relied on crash counts, bug counts, and/or
code-coverage profiles for measuring and comparing fuzzer performance. While the
problems with crash counts and code-coverage profiles are well known (see
\autoref{sec:crash-count}), in our view, simply counting the number of bugs
discovered is too coarse-grained. Instead, we argue that it is important to
distinguish between \emph{reaching}, \emph{triggering}, and \emph{detecting} a
bug. Consequently, \sys uses these three bug-centric performance metrics to
evaluate fuzzers.

A \emph{reached} bug refers to a bug whose oracle was called, implying that the
executed path reaches the context of the bug, without necessarily triggering a
fault. This is where coverage profiles fall short: simply covering the faulty
code does not mean that the program is in the correct state to trigger the bug.
Hence, a \emph{triggered} bug refers to a bug that was reached, and \emph{whose
triggering condition was satisfied}, indicating that a fault occurred. Whereas
triggering a bug implies that the program has transitioned into a faulty state,
the symptoms of the fault may not be directly observable at the oracle injection
site. When a bug is triggered, the oracle only indicates that the conditions for
a fault have been satisfied, but this does not imply that the fault was
encountered or detected by the fuzzer.

Source-code instrumentation (i.e., the canary) provides ground-truth knowledge
and runtime feedback of reached and triggered bugs. Each bug is approximated by
\begin{inlinealph}
\item the lines of code patched in response to a bug report, and

\item a boolean expression representing the bug's trigger condition.
\end{inlinealph}
The canary reports:
\begin{inlineroman}
\item when the line of code is reached; and

\item when the input satisfies the conditions for faulty behavior (i.e.,
triggers the bug).
\end{inlineroman}
\autoref{sec:leaky-oracles} discusses how we prevent canaries from leaking
information to the system-under-test.

Finally, we also draw a distinction between \emph{triggering} and
\emph{detecting} a bug. Whereas most security-critical bugs manifest as a
low-level security policy violation for which state-of-the-art sanitizers are
well-suited (e.g., memory corruption, data races, invalid arithmetic), other bug
classes are not as easily observed. For example, resource exhaustion bugs are
often detected long after the fault has manifested, either through a timeout or
an out-of-memory error. Even more obscure are semantic bugs, whose malfunctions
cannot be observed without a specification or reference. Consequently, various
fuzzing techniques have been developed to target these bug classes (e.g.,
SlowFuzz~\cite{slowfuzz} and NEZHA~\cite{nezha}). Such advancements in fuzzer
techniques may benefit from an evaluation which includes the bug
\emph{detection} rate as another dimension for comparison.

\subsection{Runtime Monitoring}
\label{sec:runtime-monitor}

\sys provides a runtime monitor that collects real-time statistics from the
instrumented target. This provides a mechanism for visualizing the fuzzer's
progress and its evolution over time, without complicating the instrumentation.

The runtime monitor collects data about reached and triggered bugs
(\autoref{sec:performance-metrics}). Because this data primarily relates to the
fuzzer's program exploration capabilities, we post-process the monitor's output
to study the fuzzer's fault detection capabilities. This is achieved by
replaying the crashing inputs (produced by the fuzzer) against the benchmark
canaries to determine which bugs were triggered and hence detected. Importantly,
it is possible that the fuzzer produces crashing inputs that do not correspond
to any injected bug. If this occurs, the new bug is triaged and added to the
benchmark for other fuzzers to discover.

\section{Design and Implementation Decisions}
\label{sec:design}

\sys's unapologetic focus on fuzzing (as opposed to being a general
bug-detection benchmark) necessitates a number of key design and implementation
choices. We discuss these choices here.

\subsection{Forward-Porting}
\label{sec:forward-porting}

\subsubsection{Forward-Porting vs.\ Back-Porting}
\label{sec:foward-vs-back-porting}

In contrast to back-porting bugs to previous versions,
forward-porting ensures that all \emph{known} bugs are fixed, and that the
reintroduced bugs will have ground-truth oracles. While it is possible that the
new fixes and features in newer codebases may (re)introduce unknown bugs,
forward-porting allows \sys to evolve with each published bug fix. Additionally,
future code changes may render a forward-ported bug obsolete, or make its
trigger conditions unsatisfiable. Without verification, forward-porting may
inject bugs which cannot be triggered. We use fuzzing to reduce this
possibility, reducing the cost of manually verifying injected bugs. A
fuzzer-generated PoV demonstrates that the bug is triggerable. Bugs that are
discovered this way are added to the list of verified bugs, helping the
evaluation of other fuzzers. While this approach may skew \sys towards
fuzzer-discoverable bugs, we argue that this is a nonissue: any newly-discovered
PoV will update the benchmark, thus ensuring a fair and balanced bug
distribution.

\subsubsection{Manual Forward-Porting}
\label{sec:manual-forward-porting}

All \sys bugs are manually introduced. This process
involves:
\begin{inlineroman}
\item searching for bug reports;
\item identifying bugs that affect the core codebase;
\item finding the relevant fix commits;
\item recognizing the bug conditions from the fix commits;
\item collecting these conditions as a set of path constraints;
\item modeling these path constraints as a boolean expression (the bug canary);
and
\item injecting these canaries to flag bugs at runtime.
\end{inlineroman}
The complexity of this process led us to reject a wholly-automated approach;
automating bug injection would likely result in an incomplete and error-prone
technique, ultimately yielding fewer bugs of lower quality. Moreover, an
automated approach still requires manual verification of the results. Dedicating
human resources to the forward-porting process maximizes the correctness of
\sys's bugs.

To justify a manual approach, we enumerate the \emph{scopes} (i.e., code blocks,
functions, modules) spanned by each bug fix and use these scopes as a measure of
bug-porting complexity (scope measures for all bugs are given in
\autoref{tab:all-bugs}). While a simple bug-porting technique works well for
fixes with a scope of one, the bug-porting technique must become more advanced
as the number of scopes increases (e.g., it must handle \emph{interprocedural}
constraints). Of the \sysbugcount \sys bugs, \SI{34}{\percent} had a scope
measure greater than one.

Finally, our manual porting process was heavily reliant on prose; in particular,
by the comments and discussions contained within bug reports. These discussions
provide valuable insight into
\begin{inlinealph}
\item developers' intent, and
\item the construction of precise trigger conditions.
\end{inlinealph}
Additionally, function names (particularly those from the standard library)
provide key insight into the code's objective, without requiring in-depth
analysis into what each function does. An automated technique would require
either:
\begin{inlineroman}
\item an in-depth analysis of such functions, likely resulting in path
explosion; or

\item inference of bug conditions and function utilities via natural language
processing (NLP).
\end{inlineroman}
Both of these approaches are too complex to be included in the scope of \sys's
development and would likely require several years of research to be effective.

\begin{figure}[b]
\begin{lstlisting}[language=C, label=lst:weird-states, escapeinside={(*}{*)},
xleftmargin=.2\textwidth, xrightmargin=.2\textwidth,
caption=Weird states can result in execution traces which do not exist in the
context of normal program behavior.]
void libfoo_baz(char *str) {
  struct { char buf[16]; size_t len; } tmp;
  tmp.len = strlen(str);
  // Bug 1: possible OOB write in strcpy()
  magma_log(1, tmp.len >= sizeof(tmp.buf));(* \label{line:bug1_capture} *)
  strcpy(tmp.buf, str);(* \label{line:bug1_trigger}*)
  // Bug 2: possible div-by-zero if tmp.len == 0
  magma_log(2, tmp.len == 0);(* \label{line:bug2_capture} *)
  int repeat = 64 / tmp.len;(* \label{line:bug2_trigger} *)
  int padlen = 64 %
}
\end{lstlisting}
\end{figure}

\subsection{Weird States}
\label{sec:weird-states}

When a fuzzer generates an input that triggers an undetected bug, and execution
continues past this bug, the program transitions into an undefined state: a
\emph{weird state}~\cite{weirdmachines}. Any information collected after
transitioning to a weird state is unreliable. To address this issue, we allow
the fuzzer to continue the execution trace, but only collect bug oracle data
\emph{before and until} the first bug is triggered (i.e., transition to a weird
state). Oracles do not signify that a bug has been executed; they only indicate
whether the conditions required to execute a bug are satisfied.

\autoref{lst:weird-states} shows an example of the interplay between weird
states. This example contains two bugs: an out-of-bounds write (bug~1) and a
division-by-zero (bug~2). When \texttt{tmp.len~==~0}, the condition for bug~1
(line~\autoref{line:bug1_trigger}) remains unsatisfied, logging and triggering
bug~2 instead (lines~\autoref{line:bug2_capture}
and~\autoref{line:bug2_trigger}, respectively). However, when
\texttt{tmp.len~>~16}, bug~1 is logged and triggered
(lines~\autoref{line:bug1_capture} and \autoref{line:bug1_trigger},
respectively). Furthermore, \texttt{tmp.len} is overwritten by a non-zero value,
leaving bug~2 untriggered. In contrast, bug~1 is triggered when
\texttt{tmp.len~==~16}, overwriting \texttt{tmp.len} with the \texttt{NULL}
terminator and setting its value to~0 (on a Little-Endian system). This also triggers
bug~2, despite the input not explicitly specifying a zero-length \texttt{str}.

\subsection{A Static Benchmark}
\label{sec:benchmark-tuned-fuzzers}

Much like other widely-used performance benchmarks---e.g., SPEC
CPU~\cite{specbms} and DaCapo~\cite{dacapo}---\sys is a \emph{static} benchmark
that contains realistic workloads. These benchmarks assume that if the
system-under-test performs well on the benchmark's workloads, then it will
perform similarly on real workloads. While realistic, static benchmarks are
susceptible to \emph{overfitting}. Overfitting can occur if developers tweak the
system-under-test to perform better on a benchmark, rather than focusing on real
workloads.

Overfitting could be overcome by \emph{dynamically synthesizing} a benchmark
(and ensuring that the system-under-test is unaware of the synthesis
parameters). However, this approach risks generating workloads different from
real-world scenarios, rendering the evaluation biased and/or incomplete. While
program synthesis is a well-studied topic~\cite{progsynth, fudge, fuzzgen}, it
remains difficult to generate large programs that remain faithful to real
development patterns and styles.

To prevent overfitting, \sys's forward-porting process allows targets to be
updated as they evolve in the real-world. Each forward-ported bug requires
minimal code changes: the addition of \sys's instrumentation and the faulty
code itself. This makes it relatively straightforward to update targets,
including introducing new bugs and new features. For example, two undergraduate
students without software security experience added over~$60$ bugs in three new
targets over a single semester.
These measures ensure that \sys remains representative of real, complex targets
and suitable for fuzzer evaluation.

\subsection{Leaky Oracles}
\label{sec:leaky-oracles}

Introducing oracles into the benchmark may leak information that interferes with
a fuzzer's exploration capability, potentially leading to overfitting (as
discussed in \autoref{sec:benchmark-tuned-fuzzers}). For example, if oracles
were implemented as \texttt{if}~statements, fuzzers that maximize branch
coverage could detect the oracle's branch and hence generate an input that
satisifies the branch condition.

One possible solution to this \emph{leaky oracle} problem is to produce both
instrumented and uninstrumented target binaries (with respect to \sys's
instrumentation, not any instrumentation that the fuzzer injects). The fuzzer's
input would be fed into both binaries, but the fuzzer would only collect the
data it needs (e.g., coverage feedback) from the uninstrumented binary. The
instrumented binary would collect canary data and report it to the runtime
monitor. This approach, however, introduces other challenges associated with
duplicating the execution trace between two binaries (e.g., replicating the
environment, maintaining synchronization between executions), greatly
complicating \sys's implementation and introducing runtime overheads.

Instead, we use \emph{always-evaluate memory writes}, whereby an injected bug
oracle evaluates a boolean expression representing the bug's trigger condition.
This typically involves a binary comparison operator, which most compilers
(e.g., gcc, clang) translate into a pair of \texttt{cmp} and \texttt{set}
instructions embedded into the execution path. The results of this evaluation
are then shared with the runtime monitor (\autoref{sec:runtime-monitor}). This
process is demonstrated in Listings~\ref{lst:canary-inst}
and~\ref{lst:canary-example}.

\begin{figure}[t]
\centering

\begin{minipage}[t]{0.52\linewidth}
\begin{lstlisting}[language=C, caption=\sys instrumentation.,
                   label=lst:canary-inst, escapeinside={(*}{*)}]
void magma_log(int id, bool condition) {
  extern struct magma_bug *bugs; // = mmap(...) (* \label{line:canary_mmap} *)
  extern bool faulty; // = false initially
  bugs[id].reached   += 1         & (faulty ^ 1); (* \label{line:canary_reached} *)
  bugs[id].triggered += condition & (faulty ^ 1); (* \label{line:canary_triggered} *)
  faulty = faulty | condition; (* \label{line:canary_faulty} *)
}
\end{lstlisting}
\end{minipage}\hfill
\begin{minipage}[t]{0.4\linewidth}
\begin{lstlisting}[language=C, caption=Instrumented example.,
                   label=lst:canary-example, escapeinside={(*}{*)}]
void libfoo_bar() {
  // uint32_t a, b, c;
  magma_log(42, (a == 0) | (b == 0)); (* \label{line:magma_log} *)
  // possible divide-by-zero
  uint32_t x = c / (a * b); (* \label{line:canary_bug} *)
}
\end{lstlisting}
\end{minipage}
\end{figure}

\autoref{lst:canary-inst} shows \sys's canary implementation. The
always-evaluated memory accesses are shown on
lines~\autoref{line:canary_reached} and~\autoref{line:canary_triggered}. The
\texttt{faulty} flag addresses the problem of weird states
(\autoref{sec:weird-states}), and disables future canaries after the first bug
is encountered.

\autoref{lst:canary-example} shows an example program instrumented with a
canary. A call to \texttt{magma\_log} is inserted
(line~\autoref{line:magma_log}) prior to the execution of the faulty code
(line~\autoref{line:canary_bug}). Compound trigger conditions---i.e., those
including the logical \texttt{and} and \texttt{or} operators---often generate
implicit branches at compile-time (due to short-circuit compiler behavior). To
avoid leaking information through coverage, we provide custom x86-64 assembly
blocks to evaluate these logical operators in a single basic block (without
short-circuit behavior). We revert to C's bitwise operators
(\texttt{\&}~and~\texttt{|})---which are more brittle and susceptible to
safety-agnostic compiler passes~\cite{llvmsidechannel}---when the compilation
target is not~x86-64.

Although this approach may introduce memory access patterns that are detectable
by taint tracking and other data-flow analysis techniques, statistical tests can
be used to infer whether the fuzzer overfits to these access patterns. By
repeating the fuzzing campaign with the uninstrumented binary, we can verify if
the results vary significantly.

\subsection{Proofs of Vulnerability}

In order to increase confidence in the injected bugs, a proof of vulnerability
(PoV) input must be supplied for every bug, verifying that the bug can be
triggered. The process of manually crafting PoVs, however, is arduous and
requires domain-specific knowledge, both about the input format and the target
program, potentially bringing the bug-injection process to a grinding halt.

When available, we extract PoVs from public bug reports. When no PoV is
available, we launch multiple fuzzing campaigns against these targets in an
attempt to trigger each injected bug. Inputs that trigger a bug are saved as a
PoV. Bugs which are not triggered, even after multiple campaigns, are manually
inspected to verify path reachability and satisfiability of trigger conditions.

\subsection{Unknown Bugs}

Because \sys uses real-world programs, it is possible that bugs exist for which
no ground-truth is available (i.e., an oracle does not exist). A fuzzer might
inadvertantly trigger these bugs and (correctly) detect a fault. Due to the
imperfections in automated deduplication techniques, these crashes are not
included in \sys's metrics. Instead, such crashes are used to improve \sys
itself. The bug's root cause can be determined by manually studying the
execution trace, after which the bug can be added to the benchmark.

\subsection{Fuzzer Compatibility}

Fuzzers are not limited to a specific execution engine under which they analyze
and explore a program. For example, some fuzzers (e.g.,
Driller~\cite{driller}, T-Fuzz~\cite{tfuzz}) leverage symbolic execution (using
an engine such as angr~\cite{angr}) to explore the target. This can introduce
\begin{inlinealph}
\item incompatibilities with \sys's instrumentation, and
\item inconsistencies in the runtime environment (depending on how the symbolic
execution engine models the environment).
\end{inlinealph}

However, the defining trait of most fuzzers, in contrast to other types of
bug-finding tools, is that they concretely execute the target on the host
system. Unlike benchmarks such as the CGC and BugBench---which aim to evaluate
\emph{all} bug-finding tools---\sys is unapologetically a \emph{fuzzing}
benchmark. This includes whitebox fuzzers that use symbolic execution to guide
input generation, provided that the target is executed on the host system
(\symcc~\cite{symcc} is one such fuzzer that we include in our evaluation).

We therefore impose the following restriction on the fuzzers evaluated by \sys:
the fuzzer must execute the target in the context of an OS process, with
unrestricted access to OS facilities (e.g., system calls, libraries, file
system). This allows \sys's runtime monitor to extract canary statistics using
the operating system's services at relatively low overhead/complexity.

\section{Evaluation}
\label{sec:evaluation}

\subsection{Methodology}

We evaluated several fuzzers in order to establish the versatility of our
metrics and benchmark suite. We chose a set of \sysevalcount \emph{mutational
fuzzers} whose source code was available at the time of writing: AFL~\cite{afl},
AFLFast~\cite{aflfast}, AFL++~\cite{aflplusplus}, \fairfuzz~\cite{fairfuzz},
\mopt-AFL~\cite{moptafl}, honggfuzz~\cite{honggfuzz}, and
\symcc-AFL~\cite{symcc}. These \sysevalcount fuzzers were evaluated over
\sysevalruns identical \sysevalduration and \sysevallongduration fuzzing
campaigns for each fuzzer/target combination. This amounts to~\systotalcpuhours
of fuzzing.

To ensure fairness, benchmark parameters were identical across all fuzzing
campaigns. Each fuzzer was bootstrapped with the same set of seed files (sourced
from the original target codebase) and configured with the same timeout and
memory limits. \sys's monitoring utility was configured to poll canary
information every five seconds, and \emph{fatal canaries} mode
(\autoref{sec:overview}) was used to evaluate a fuzzer's ability to \emph{reach}
and \emph{trigger} bugs.
All experiments were run on one of three machines, each with an
Intel\textsuperscript{\textregistered} Xeon\textsuperscript{\textregistered}
Gold 5218 CPU and 64~GB of RAM, running Ubuntu 18.04 LTS 64-bit. The targets
were compiled for x86-64.

\emph{AddressSanitizer} (ASan)~\cite{asan} was used to evaluate \emph{detected}
bugs. Crashing inputs (generated by fatal canaries) were validated by replaying
them through the ASan-instrumented target. Although this evaluation method
measures ASan's fault-detection capabilities, it still highlights the bugs that
fuzzers can realistically detect when fuzzing without ground truth.

\subsection{Time to Bug}

We use the time required to find a bug as a measure of fuzzer performance. As discussed
in \autoref{sec:performance-metrics}, \sys records the time taken to both reach
and trigger a bug, allowing us to compare fuzzer performance across multiple
dimensions. Fuzzing campaigns are typically limited to a finite
duration (we limit our campaigns to~\sysevalduration and \sysevallongduration,
repeated \sysevalruns times), so it is important that the time-to-bug discovery
is low.

The highly-stochastic nature of fuzzing means that the time-to-bug can vary
wildly between identical trials. To account for this variation, we repeat each
trial \sysevalruns times. Despite this repetition, a fuzzer may still fail to
find a bug within the alloted time, leading to missing measurements. We
therefore apply \emph{survival analysis} to account for this missing data and
high variation in bug discovery times. Specifically, we adopt Wagner's
approach~\cite{wagnerphd} and use the Kaplan-Meier estimator~\cite{kaplanmeier}
to model a bug's \emph{survival function}. This survival function describes the
probability that a bug remains undiscovered (i.e., ``survives'') within a given
time (here, \sysevalduration and \sysevallongduration trials). A smaller
survival time indicates better fuzzer performance.

\subsection{Experimental Results}

\autoref{fig:distributions}, \autoref{fig:signplot}, \autoref{table:eval}, and
\autoref{table:eval_7d} present the results of our fuzzing campaigns.

\subsubsection{Bug Count and Statistical Significance}

\begin{figure}
\centering
\resizebox{0.75\textwidth}{!}{\includegraphics{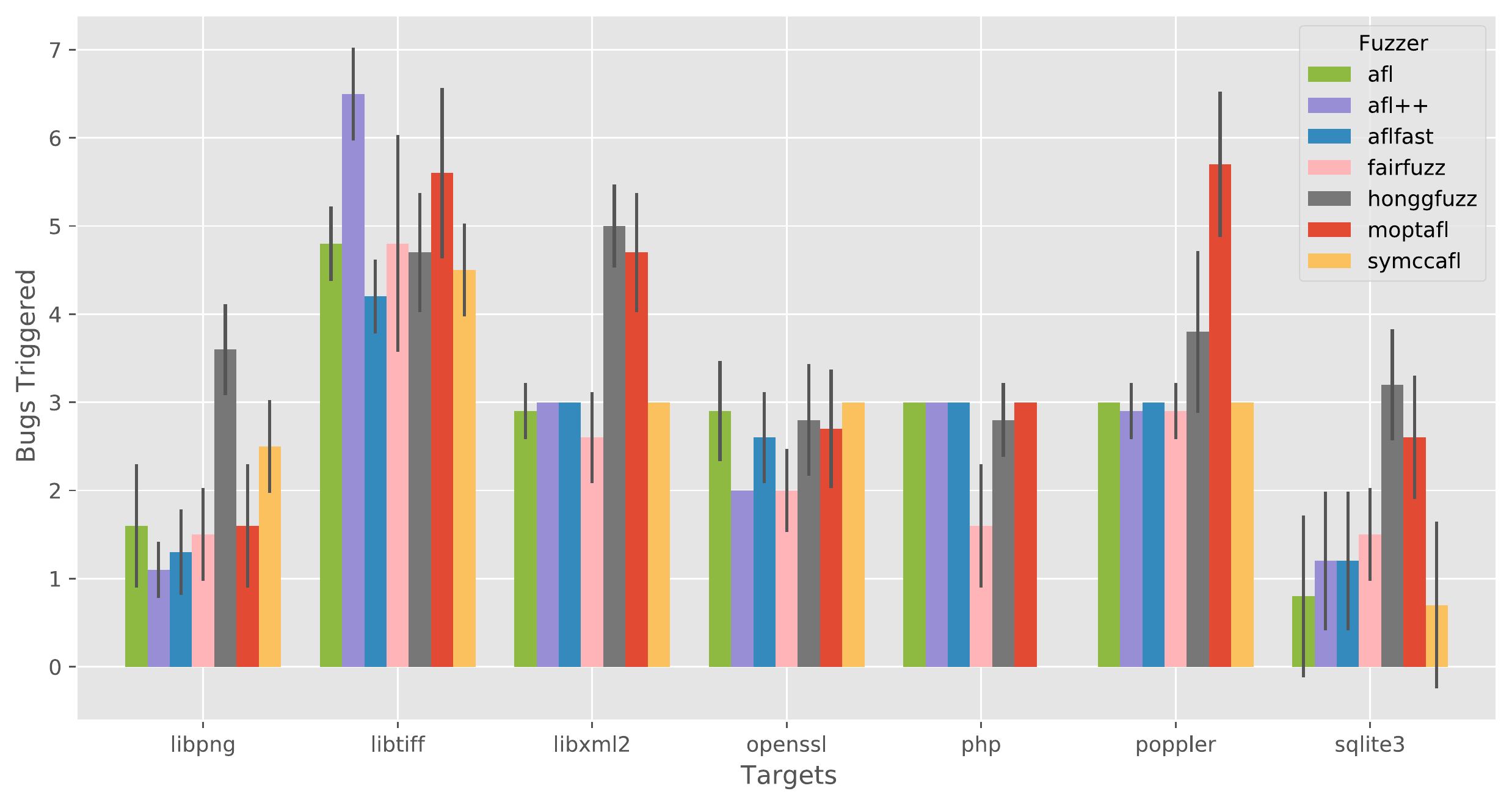}}
\captionof{figure}{The mean number of bugs (and standard deviation) found by
each fuzzer across \sysevalruns \sysevalduration campaigns.}
\label{fig:distributions}
\end{figure}

\begin{figure}
\centering
\resizebox{\textwidth}{!}{\includegraphics{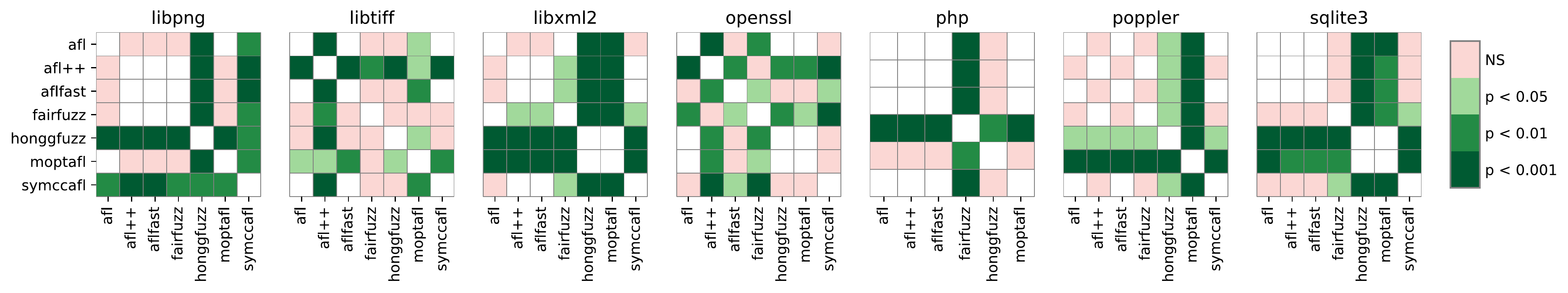}}
\captionof{figure}{Significance of evaluations of fuzzer pairs using p-values
from the Mann-Whitney U-Test. We use $p<0.05$ as a threshold for significance.
Values greater than~$0.05$ are shaded red. Darker shading indicates a lower
p-value, or higher statistical significance. White cells indicate that the pair
of sample sets are identical.}
\label{fig:signplot}
\end{figure}

\autoref{fig:distributions} shows the mean number of bugs found per fuzzer
(across \sysevalruns \sysevalduration campaigns).  These values are susceptible
to outliers, limiting the conclusions that we can draw about fuzzer performance.
We therefore conducted a statistical significance analysis of the collected
sample-set pairs to calculate p-values using the Mann-Whitney U-test.  P-values
provide a measure of how different a pair of sample sets are, and how
significant these differences are. Because our results are collected from
independent populations (i.e., different fuzzers), we make no assumptions about
their distributions. Hence, we apply the Mann-Whitney U-test to measure
statistical significance. \autoref{fig:signplot} shows the results of this
analysis.

The Mann-Whitney U-test shows that AFL, AFLFast, AFL++, and \symcc-AFL performed
similarly against most targets (signified by the large number of red and white
cells in \autoref{fig:signplot}), despite some minor differences in mean bug
counts (shown in \autoref{fig:distributions}). \autoref{fig:signplot} shows
that, in most cases, the small fluctuations in mean bug counts are not
significant, and the results are thus not sufficiently conclusive. One oddity is
the performance of AFL++ against \emph{libtiff}. \autoref{fig:distributions}
reveals that AFL++ scored the highest mean bug count compared to all other
fuzzers, and \autoref{fig:signplot} shows that this difference is statistically
significant.

On the other hand, \fairfuzz~\cite{fairfuzz} displayed significant performance
regression against \emph{libxml2}, \emph{openssl}, and \emph{php}. While the
original evaluation of \fairfuzz claims that it achieved the highest coverage
against \texttt{xmllint}, that improvement was not reflected in our results.

Finally, honggfuzz and \mopt-AFL performed significantly better than all other
fuzzers in three out of \systargetcount targets. Additionally, honggfuzz was the
best fuzzer for \textit{libpng} as well. We attribute honggfuzz's performance to its wrapping of
memory-comparison functions, which provides comparison progress information to
the fuzzer (similar to Steelix~\cite{steelix}).

\subsubsection{Time to Bug}

In total, during the \sysevalduration campaigns, \sysreachbugcount of
the~\sysbugcount \sys bugs (\SI{62}{\percent}) were reached.
Additionally,~\systriggerbugcount of the~\syspovcount \emph{verified} bugs
(\SI{79}{\percent})---i.e., those with PoVs---were triggered. Notably, no single
fuzzer triggered more than~37 bugs (\SI{68}{\percent} of the verified bugs).
These results are presented in \autoref{table:eval}. Here, bugs are sorted by
the mean trigger time, which we use to approximate ``difficulty''.

The long bug discovery times (19 of the~\systriggerbugcount triggered
bugs---\SI{44}{\percent}---took on average more than \SI{20}{\hour} to trigger)
suggests that the evaluated fuzzers still have a long way to go in improving
program exploration. However, while many of the \sys bugs are difficult to
discover, \autoref{table:eval} highlights a set of~17 ``simple'' bugs that all
fuzzers find consistently within~\sysevalduration. These bugs provide a baseline
for detecting performance regression: if a new fuzzer fails to discover these
bugs, then its program exploration strategy should be revisited.

Most of the bugs in \autoref{table:eval} were reached by all fuzzers. \symcc-AFL
was the worst performing fuzzer in this regard, failing to reach nine bugs (the
highest amongst the~\sysevalcount evaluated fuzzers).
Interestingly, most bugs show a large difference between reach and trigger
times. For example, only the first three bugs listed in \autoref{table:eval}
were triggered when first reached. In contrast, bugs such as MAE115 (from
\textit{openssl}) take \SI{10}{\second} to reach (by all fuzzers), but up to
\SI{20}{\hour} (on average) to trigger. This difference between time-to-reach
and time-to-trigger a bug provides another feature for determining bug
``difficulty'': while control flow may be trivially satisfied (as evidence by
the time to reach a bug), bugs such as MAE115 may require complex, stateful
data-flow constraints.

The longer, \sysevallongduration campaigns in \autoref{table:eval_7d} reveal a
peculiar result: while honggfuzz was faster to trigger bugs during the
\sysevalduration campaigns, \mopt-AFL was faster to trigger~11 additional bugs
after \sysevalduration, making it the most successful fuzzer over the
\sysevallongduration campaigns. Notably, honggfuzz failed to trigger any of
these~11 bugs. This highlights the importance of long fuzzing campaigns and the
utility of \sys's survival time analysis for comparing fuzzer performance.

\begin{figure}[t]
  \centering
  \includegraphics[width=\linewidth]{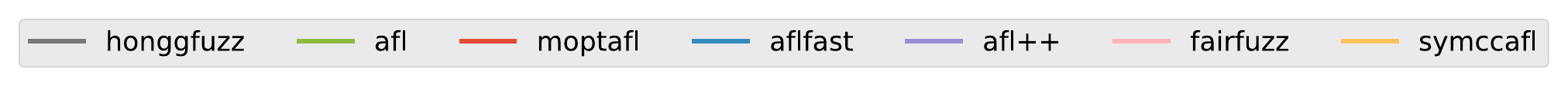}

  \begin{subfigure}[t]{0.49\linewidth}
    \centering
    \includegraphics[width=0.9\linewidth]{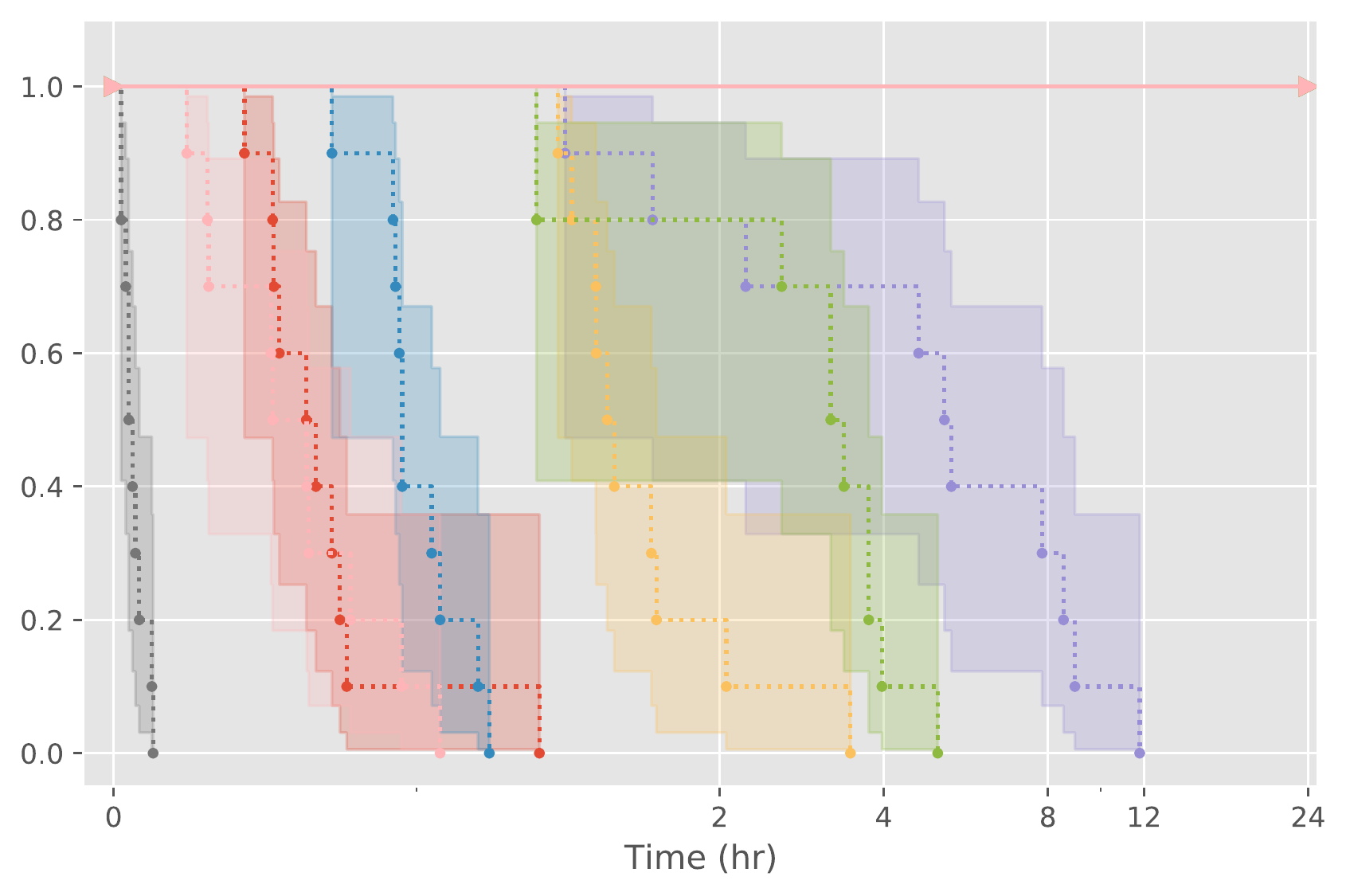}

    \caption{Bug AAH018 (\textit{libtiff} with \texttt{read\_rgba\_fuzzer}).}
    \label{fig:aah018-survival}
  \end{subfigure}\hfill
  \begin{subfigure}[t]{0.49\linewidth}
    \centering
    \includegraphics[width=0.9\linewidth]{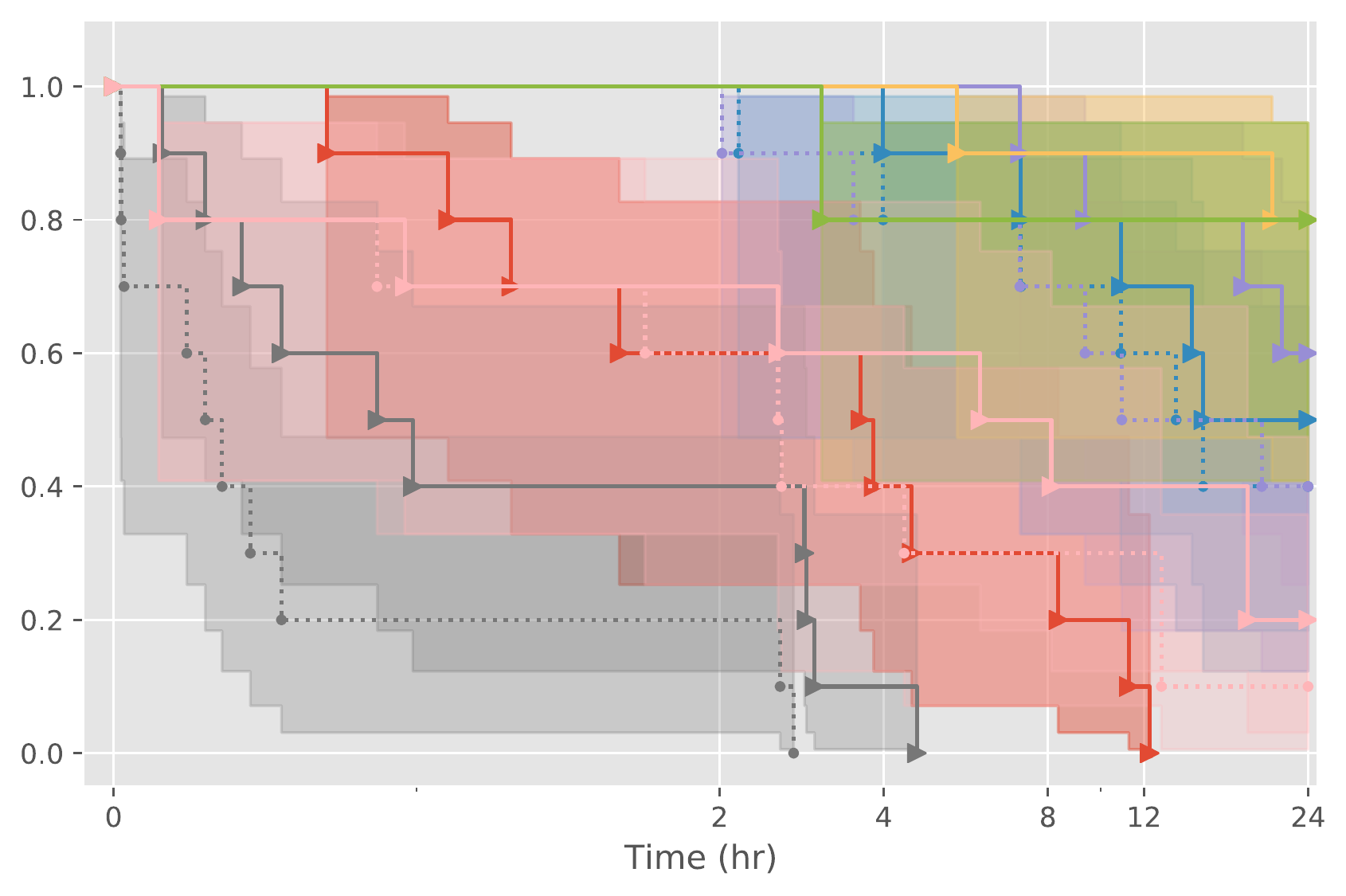}

    \caption{Bug JCH232 (\textit{sqlite3} with \texttt{sqlite3\_fuzz}).}
    \label{fig:jch232-survival}
  \end{subfigure}

  \begin{subfigure}[t]{0.49\linewidth}
    \centering
    \includegraphics[width=0.9\linewidth]{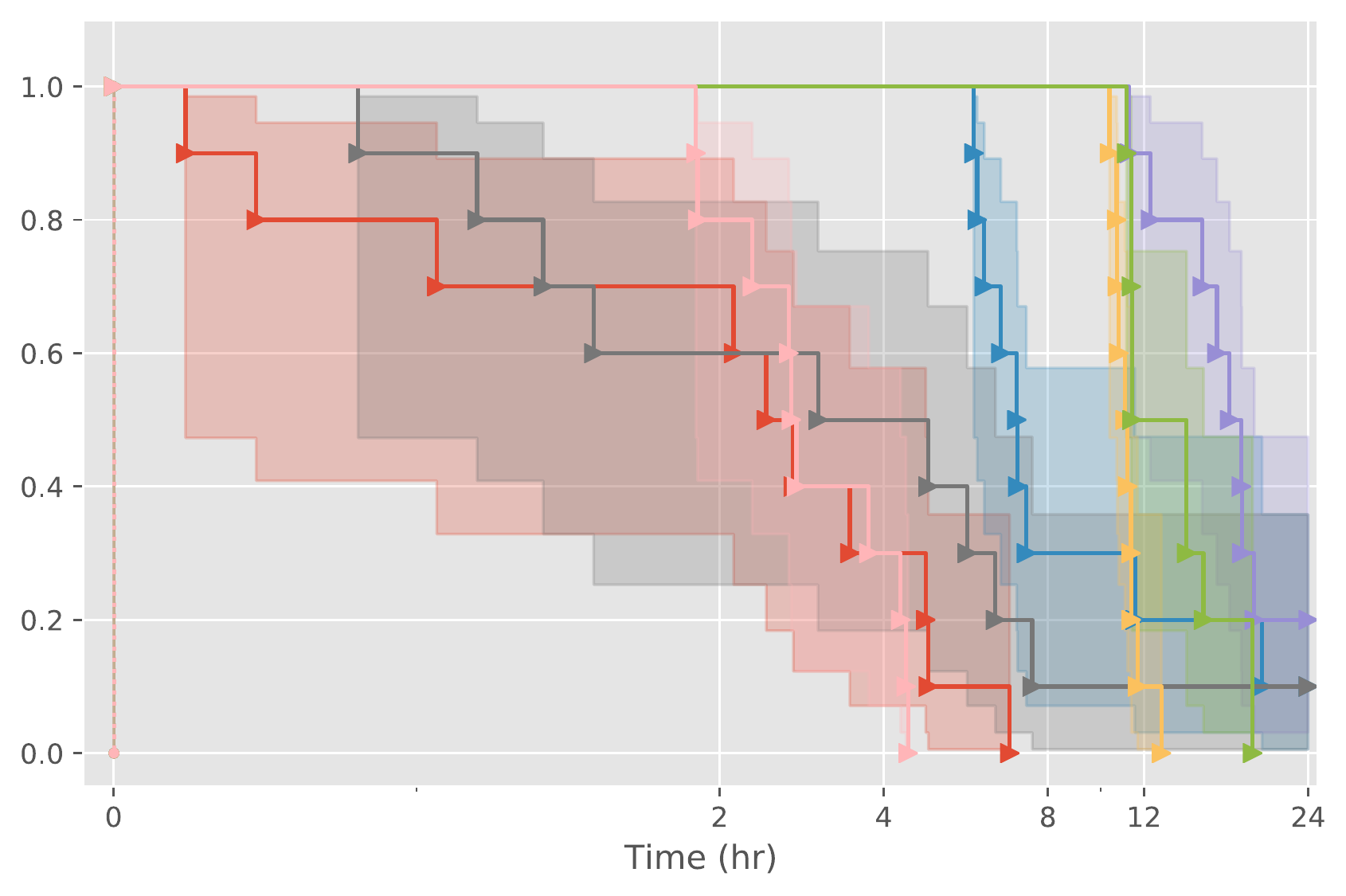}

    \caption{Bug AAH020 (\textit{libtiff} with \texttt{tiffcp}).}
    \label{fig:aah020-survival-2}
  \end{subfigure}\hfill
  \begin{subfigure}[t]{0.49\linewidth}
    \centering
    \includegraphics[width=0.9\linewidth]{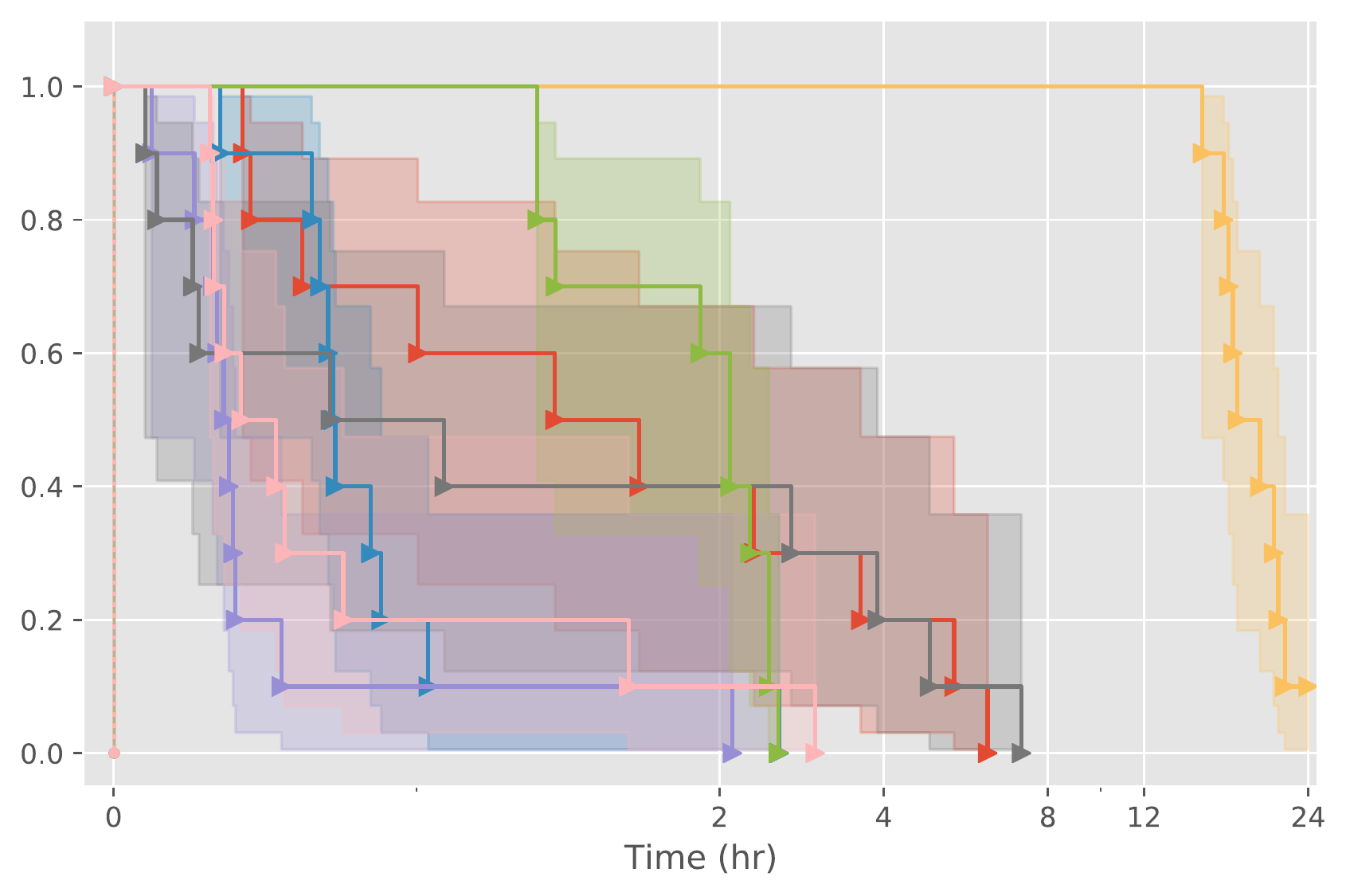}

    \caption{Bug AAH020 (\textit{libtiff} with \texttt{read\_rgba\_fuzzer}).}
    \label{fig:aah020-survival-1}
  \end{subfigure}
  \hfill

  \caption{Survival functions for a subset of \sys bugs. The $y$-axis is the
  \emph{survival probability} for the given bug. Dotted lines represent survival
  functions for \emph{reached} bugs, while solid lines represent survival
  functions for \emph{triggered} bugs. Confidence intervals are shown as shaded
  regions.}
  \label{fig:surival-funcs}
\end{figure}

\autoref{fig:surival-funcs} plots four survival functions for three \sys bugs
(AAH018, JCH232, and AAH020). These plots illustrate the probability of a bug
surviving a~\sysevalduration fuzzing trial, and are generated by applying the
Kaplan-Meier estimator to the results of \sysevalruns repeated fuzzing trials.
Dotted lines represent survival functions for \emph{reached} bugs, while solid
lines represent survival functions for \emph{triggered} bugs. Confidence
intervals are shown as shaded regions.
\autoref{fig:aah018-survival} shows the time to reach bug AAH018
(\textit{libtiff}). Notably, this bug was not triggered by any of the
\sysevalcount evaluated fuzzers. Thus, the probability of bug AAH018
``surviving'' \sysevalduration (i.e., not being triggered) remains at one.
In comparison, \autoref{fig:jch232-survival} shows the differences in the time
taken to reach and trigger bug JCH232 (\textit{sqlite3}). Here, honggfuzz is the
best performer, because the bug's probability of survival approaches zero the
fastest. Notably, the variance is much higher compared to bug AAH018 (as evident
by the larger confidence intervals).
Finally, \autoref{fig:aah020-survival-1} and \autoref{fig:aah020-survival-2}
compare the probability of survival for bug AAH020 (\textit{libtiff}) across two
driver programs: \texttt{tiffcp} and \texttt{read\_rgba\_fuzzer}. The former is
a general-purpose application, while the latter is a driver specifically
designed as a fuzzer harness. While the bug is reached relatively quickly by
both drivers, the fuzzer harness is clearly superior at \emph{triggering} the
bug, as it is faster across \emph{all} fuzzers. This result supports our claim
in \autoref{sec:target-selection} that \emph{domain experts are most suitable for
selecting and developing fuzzing drivers}.

Again, it is clear that honggfuzz outperforms all other fuzzers (in both
reaching and triggering bugs), finding~11 additional bugs not triggered by other
fuzzers. In addition to its finer-grained instrumentation, honggfuzz natively
supports persistent fuzzing. Our experiments show that honggfuzz's execution
rate was at least three times higher than that of AFL-based fuzzers using
persistent drivers. This undoubtedly contributes to honggfuzz's strong
performance.

\begin{figure}[h]
\begin{lstlisting}[language=C, label=lst:complexbug,
xleftmargin=.27\textwidth, xrightmargin=.27\textwidth,
caption=Divide-by-zero bug in \textit{libpng}. Input undergoes non-trivial
transformations to trigger the bug.]
void png_check_chunk_length(png_ptr, length) {
  size_t row_factor = png_ptr->width // uint32_t
    * png_ptr->channels // uint32_t
    * (png_ptr->bit_depth > 8? 2: 1)
    + 1
    + (png_ptr->interlaced? 6: 0);

  if (png_ptr->height > UINT_32_MAX/row_factor) {
    idat_limit = UINT_31_MAX;
  }
}
\end{lstlisting}
\end{figure}

\subsubsection{Achilles' Heel of Mutational Fuzzing}

AAH001 (CVE-2018-13785, shown in \autoref{lst:complexbug}), is a divide-by-zero bug in
\emph{libpng}. It is triggered when the input is a non-interlaced 8-bit RGB
image with a width of \texttt{0x55555555}. This ``magic value'' is not encoded
anywhere in the target, and is easily calculated by solving the constraints for
\lstinline[basicstyle=\normalsize\ttfamily]{row_factor == 0}. However,
mutational fuzzers struggle to discover this bug type. This is because
mutational fuzzers sample from an extremely large input space, making them
unlikely to pick the exact byte sequence required to trigger the bug (here,
\texttt{0x55555555}). Notably, only honggfuzz, AFL, and \symcc-AFL were able to
trigger this bug. \symcc-AFL was the fastest to do so, likely due to its
constraint-solving capabilities.

\subsubsection{Magic Value Identification}

AAH007 is a dangling pointer bug in \emph{libpng}, and illustrates how some
fuzzer features improve bug-finding ability. To trigger this bug, it is
sufficient for a fuzzer to provide a valid input with an \texttt{eXIF} chunk
(which remains unmarked for release upon object destruction, leading to a
dangling pointer). Unlike the AFL-based fuzzers, honggfuzz is able to
consistently trigger this bug relatively early in each campaign. We posit that
this is due to honggfuzz replacing the \texttt{strcmp} function with an
instrumented wrapper that incrementally satisfies string magic-value checks.
\symcc-AFL also consistently triggers this bug, demonstrating how whitebox
fuzzers can trivially solve constraints based on magic values.

\subsubsection{Semantic Bug Detection}

AAH003 (CVE-2015-8472) is a data inconsistency in \texttt{libpng}'s API, where
two references to the same piece of information (color-map size) can yield
different values. Such a semantic bug does not produce observable behavior that
violates a known security policy, and it cannot be detected by state-of-the-art
sanitizers without a specification of expected behavior.

Semantic bugs are not always benign. Privilege escalation and command
injection are two of the most security-critical logic bugs that are still found
in modern systems, but they remain difficult to detect with standard
sanitization techniques. This observation highlights the shortcomings of current
fault detection mechanisms and the need for more fault-oriented bug-finding
techniques (e.g., NEZHA~\cite{nezha}).

\subsubsection{Comparison to \lavam}\label{sec:lavam-evaluation}

In addition to our \sys evaluation, we also evaluate the same \sysevalcount
fuzzers against \lavam, measuring
\begin{inlinealph}
\item the overheads introduced by \lavam's bug oracles, and

\item the total number of bugs found by each fuzzer (across a~\SI{24}{\hour}
campaign, repeated~10 times per fuzzer).
\end{inlinealph}
These results---presented in \autoref{tab:lavam-bugs-overheads}---show that
\lavam's most iconic target, \textit{who}, accounts for~\SI{94.3}{\percent} of
the benchmark's bugs. This high bug count reduces the amount of functional code
(compared to benchmark instrumentation)
in the \textit{who} binary to~\SI{5.3}{\percent}, impeding a fuzzer's
exploration capabilities. Notably, we found that the evaluated fuzzers spent (on
average)~\SI{42.9}{\percent} of their time executing oracle code in \textit{who}
(this percentage is based on the final state of the fuzzing queue, and may not
represent the runtime overhead of \emph{all} code paths). Finally, the bug
counts found by each fuzzer show a clear bias towards fuzzers with magic-value
detection capabilities (due to \lavam's single, simple bug type, per
\autoref{sec:existing-benchmarks}).

\begin{table}
\centering

\caption{Overheads introduced by \lavam compared to \texttt{coreutils-8.24}.
These overheads denote increases in LLVM IR instruction counts, object file
sizes, and average runtimes when processing seeds generated from
a~\SI{24}{\hour} fuzzing campaign. The total number of unique bugs triggered
across all~10 trials/fuzzer is also shown, with the best performing fuzzer
highlighted in green.}
\label{tab:lavam-bugs-overheads}

\begin{adjustbox}{width=\linewidth}
\rowcolors{3}{rowgray}{white}
\begin{tabular}{l|r|rrr|rrrrrrr}
  \toprule
  \multirow{2}{*}{Target}                     &
  \multirow{2}{*}{Bugs}                       &
  \multicolumn{3}{c|}{Overheads (\%)}         &
  \multicolumn{7}{c}{Total bugs triggered (\#)} \\
                                              &         & LLVM IR        & Size    & Runtime  &
         afl                                  & aflfast & afl++          & moptafl & fairfuzz & honggfuzz & symccafl \\
  \midrule
  \textit{base64}                             & 44      & 107.9          & 57.2    & 9.7      &
         1                                    & 0       & \greencell 48  & 0       & 3        & 33        & 0         \\
  \textit{md5sum}                             & 57      & 60.2           & 46.1    & 9.5      &
         0                                    & 1       & \greencell 40  & 1       & 1        & 29        & 0         \\
  \textit{uniq}                               & 28      & 63.6           & 27.8    & 11.6     &
         3                                    & 0       & \greencell 29  & 1       & 0        & 13        & 3         \\
  \textit{who}                                & 2136    & 1786.7         & 2409.1  & 42.9     &
         1                                    & 1       & \greencell 819 & 1       & 1        & 750       & 1         \\
  \bottomrule
\end{tabular}
\end{adjustbox}
\end{table}

\subsection{Discussion}

\subsubsection{Ground Truth and Confidence}

Ground truth enables us to determine a crash's root cause. Unlike many existing
benchmarks, \sys provides straightforward access to ground truth. While ground
truth is available for all~\sysbugcount bugs, only~\SI{45}{\percent} of these
bugs have a PoV that demonstrate triggerability.  Importantly, only bugs with
PoVs can be used to confidently measure a fuzzer's performance. Regardless, bugs
without a PoV remain useful: any fuzzer evaluated against \sys can produce a
PoV, increasing the benchmark's utility. Widespread adoption of \sys will
increase the number of bugs with PoVs. Notably, \autoref{table:eval_7d} shows
that running the benchmark for longer indeed yields more PoVs for
previously-untriggered bugs. We leave it as an open challenge to generate PoVs
for these bugs.

\subsubsection{Beyond Crashes}

While \sys's instrumentation does not collect information about
\emph{detected} bugs (detection is a characteristic of the fuzzer, not the bug
itself), it does enable the evaluation of this metric through a post-processing
step (supported by fatal canaries).

In particular, bugs should not be restricted to crash-triggering faults. For
example, some bugs result in resource starvation (e.g., unbounded loops or
\texttt{malloc}s), privilege escalation, or undesirable outputs. Importantly,
fuzzer developers recognize the need for additional bug-detection mechanisms:
AFL has a hang timeout, and SlowFuzz searches for inputs that trigger worst-case
behavior. Excluding non-crashing bugs from an evaluation leads to an
under-approximation of real bugs. Their inclusion, however, enables better bug
detection tools. Evaluating fuzzers based on bugs \emph{reached},
\emph{triggered}, and \emph{detected} allows us to classify fuzzers and compare
different approaches along multiple dimensions (e.g., bugs reached allows for an
evaluation of path exploration, while bugs triggered and detected allows for an
evaluation of a fuzzer's constraint generation/solving capabilities). It also
allows us to identify which bug classes continue to evade state-of-the-art
sanitization techniques (and to what degree).

\subsubsection{\sys as a Lasting Benchmark}

\sys leverages software with a long history of security bugs to build an
extensible framework with ground truth knowledge. Like most benchmarks, the
widespread adoption of \sys defines its utility. Benchmarks provide a common
basis through which systems are evaluated and compared. For instance, the
community continues to use \lavam to evaluate and compare fuzzers, despite the
fact that most of its bugs have been found, and that these bugs are of a single,
synthetic type. \sys aims to provide an evaluation platform that incorporates
realistic bugs in real software.

\section{Conclusions}
\label{sec:conclusion}

\sys is an open ground-truth fuzzing benchmark that enables accurate and
consistent fuzzer evaluation and performance comparison. We designed and
implemented \sys to provide researchers with a benchmark containing \emph{real}
targets with \emph{real} bugs. We achieve this by forward-porting~\sysbugcount
bugs across \systargetcount diverse targets. However, this is only the
beginning. \sys's simple design and implementation allows it to be easily
improved, updated, and extended, making it ideal for open-source collaborative
development and contribution. Increased adoption will only strengthen \sys's
value, and thus we encourage fuzzer developers to incorporate their fuzzers into
\sys.

We evaluated \sys against \sysevalcount popular open-source mutation-based
fuzzers (AFL, AFLFast, AFL++, \fairfuzz, \mopt-AFL, honggfuzz, and \symcc-AFL).
Our evaluation shows that ground truth enables systematic comparison of fuzzer
performance. Our evaluation provides tangible insight into fuzzer performance,
why crash counts are often misleading, and how randomness affects fuzzer
performance. It also brought to light the shortcomings of some existing fault
detection methods used by fuzzers.

Despite best practices, evaluating fuzz testing remains challenging. With the
adoption of ground-truth benchmarks like \sys, fuzzer evaluation will become
reproducible, allowing researchers to showcase the true contributions of new
fuzzing approaches. \sys is open-source and available at \sysurl.

\bibliographystyle{ACM-Reference-Format}
\bibliography{ms}


\begin{thebibliography}{66}

%%% the Web address.

\ifx \showCODEN    \undefined \def \showCODEN     #1{\unskip}     \fi
\ifx \showDOI      \undefined \def \showDOI       #1{#1}\fi
\ifx \showISBNx    \undefined \def \showISBNx     #1{\unskip}     \fi
\ifx \showISBNxiii \undefined \def \showISBNxiii  #1{\unskip}     \fi
\ifx \showISSN     \undefined \def \showISSN      #1{\unskip}     \fi
\ifx \showLCCN     \undefined \def \showLCCN      #1{\unskip}     \fi
\ifx \shownote     \undefined \def \shownote      #1{#1}          \fi
\ifx \showarticletitle \undefined \def \showarticletitle #1{#1}   \fi
\ifx \showURL      \undefined \def \showURL       {\relax}        \fi
% invisible to TeX
\providecommand\bibfield[2]{#2}
\providecommand\bibinfo[2]{#2}
\providecommand\natexlab[1]{#1}
\providecommand\showeprint[2][]{arXiv:#2}

\bibitem[\protect\citeauthoryear{Afanador and Irvine}{Afanador and
  Irvine}{2020}]%
        {bvat}
\bibfield{author}{\bibinfo{person}{Kayla Afanador} {and}
  \bibinfo{person}{Cynthia Irvine}.} \bibinfo{year}{2020}\natexlab{}.
\newblock \showarticletitle{Representativeness in the Benchmark for
  Vulnerability Analysis Tools (B-VAT)}. In \bibinfo{booktitle}{\emph{13th
  {USENIX} Workshop on Cyber Security Experimentation and Test ({CSET} 20)}}.
  \bibinfo{publisher}{{USENIX} Association}.
\newblock
\urldef\tempurl%
\url{https://www.usenix.org/conference/cset20/presentation/afanador}
\showURL{%
\tempurl}


\bibitem[\protect\citeauthoryear{Aizatsky, Serebryany, Chang, Arya, and
  Whittaker}{Aizatsky et~al\mbox{.}}{2016}]%
        {ossfuzz}
\bibfield{author}{\bibinfo{person}{Mike Aizatsky}, \bibinfo{person}{Kostya
  Serebryany}, \bibinfo{person}{Oliver Chang}, \bibinfo{person}{Abhishek Arya},
  {and} \bibinfo{person}{Meredith Whittaker}.} \bibinfo{year}{2016}\natexlab{}.
\newblock \bibinfo{title}{Announcing {OSS-Fuzz}: Continuous fuzzing for open
  source software}.
\newblock
  \bibinfo{howpublished}{\url{https://opensource.googleblog.com/2016/12/announcing-oss-fuzz-continuous-fuzzing.html}}.
\newblock
\newblock
\shownote{{Accessed: 2019-09-09}.}


\bibitem[\protect\citeauthoryear{Archer and Darkkey}{Archer and
  Darkkey}{[n.d.]}]%
        {radamsa}
\bibfield{author}{\bibinfo{person}{Branden Archer} {and}
  \bibinfo{person}{Darkkey}.} \bibinfo{year}{[n.d.]}\natexlab{}.
\newblock \bibinfo{title}{{radamsa}: A Black-box mutational fuzzer}.
\newblock \bibinfo{howpublished}{\url{https://gitlab.com/akihe/radamsa}}.
\newblock
\newblock
\shownote{{Accessed: 2019-09-09}.}


\bibitem[\protect\citeauthoryear{Arkin}{Arkin}{2009}]%
        {adobefuzz}
\bibfield{author}{\bibinfo{person}{Brad Arkin}.}
  \bibinfo{year}{2009}\natexlab{}.
\newblock \bibinfo{title}{{Adobe Reader and Acrobat Security Initiative}}.
\newblock
  \bibinfo{howpublished}{\url{http://blogs.adobe.com/security/2009/05/adobe_reader_and_acrobat_secur.html}}.
\newblock
\newblock
\shownote{{Accessed: 2019-09-09}.}


\bibitem[\protect\citeauthoryear{Arya and Neckar}{Arya and Neckar}{2012}]%
        {gfuzzforsec}
\bibfield{author}{\bibinfo{person}{Abhishek Arya} {and} \bibinfo{person}{Cris
  Neckar}.} \bibinfo{year}{2012}\natexlab{}.
\newblock \bibinfo{title}{Fuzzing for security}.
\newblock
  \bibinfo{howpublished}{\url{https://blog.chromium.org/2012/04/fuzzing-for-security.html}}.
\newblock
\newblock
\shownote{{Accessed: 2019-09-09}.}


\bibitem[\protect\citeauthoryear{Babic, Bucur, Chen, Ivancic, King, Kusano,
  Lemieux, Szekeres, and Wang}{Babic et~al\mbox{.}}{2019}]%
        {fudge}
\bibfield{author}{\bibinfo{person}{Domagoj Babic}, \bibinfo{person}{Stefan
  Bucur}, \bibinfo{person}{Yaohui Chen}, \bibinfo{person}{Franjo Ivancic},
  \bibinfo{person}{Tim King}, \bibinfo{person}{Markus Kusano},
  \bibinfo{person}{Caroline Lemieux}, \bibinfo{person}{László Szekeres},
  {and} \bibinfo{person}{Wei Wang}.} \bibinfo{year}{2019}\natexlab{}.
\newblock \showarticletitle{FUDGE: Fuzz Driver Generation at Scale}. In
  \bibinfo{booktitle}{\emph{Proceedings of the 2019 27th ACM Joint Meeting on
  European Software Engineering Conference and Symposium on the Foundations of
  Software Engineering}}.
\newblock


\bibitem[\protect\citeauthoryear{Blackburn, Garner, Hoffmann, Khang, McKinley,
  Bentzur, Diwan, Feinberg, Frampton, Guyer, Hirzel, Hosking, Jump, Lee, Moss,
  Phansalkar, Stefanovi\'{c}, VanDrunen, von Dincklage, and
  Wiedermann}{Blackburn et~al\mbox{.}}{2006}]%
        {dacapo}
\bibfield{author}{\bibinfo{person}{Stephen~M. Blackburn},
  \bibinfo{person}{Robin Garner}, \bibinfo{person}{Chris Hoffmann},
  \bibinfo{person}{Asjad~M. Khang}, \bibinfo{person}{Kathryn~S. McKinley},
  \bibinfo{person}{Rotem Bentzur}, \bibinfo{person}{Amer Diwan},
  \bibinfo{person}{Daniel Feinberg}, \bibinfo{person}{Daniel Frampton},
  \bibinfo{person}{Samuel~Z. Guyer}, \bibinfo{person}{Martin Hirzel},
  \bibinfo{person}{Antony Hosking}, \bibinfo{person}{Maria Jump},
  \bibinfo{person}{Han Lee}, \bibinfo{person}{J.~Eliot~B. Moss},
  \bibinfo{person}{Aashish Phansalkar}, \bibinfo{person}{Darko Stefanovi\'{c}},
  \bibinfo{person}{Thomas VanDrunen}, \bibinfo{person}{Daniel von Dincklage},
  {and} \bibinfo{person}{Ben Wiedermann}.} \bibinfo{year}{2006}\natexlab{}.
\newblock \showarticletitle{The DaCapo Benchmarks: Java Benchmarking
  Development and Analysis}. In \bibinfo{booktitle}{\emph{Proceedings of the
  21st Annual ACM SIGPLAN Conference on Object-Oriented Programming Systems,
  Languages, and Applications}} (Portland, Oregon, USA)
  \emph{(\bibinfo{series}{OOPSLA ’06})}. \bibinfo{publisher}{Association for
  Computing Machinery}, \bibinfo{address}{New York, NY, USA},
  \bibinfo{pages}{169–190}.
\newblock
\showISBNx{1595933484}
\urldef\tempurl%
\url{https://doi.org/10.1145/1167473.1167488}
\showDOI{\tempurl}


\bibitem[\protect\citeauthoryear{Blazytko, Schl{\"o}gel, Aschermann, Abbasi,
  Frank, W{\"o}rner, and Holz}{Blazytko et~al\mbox{.}}{2020}]%
        {aurora}
\bibfield{author}{\bibinfo{person}{Tim Blazytko}, \bibinfo{person}{Moritz
  Schl{\"o}gel}, \bibinfo{person}{Cornelius Aschermann}, \bibinfo{person}{Ali
  Abbasi}, \bibinfo{person}{Joel Frank}, \bibinfo{person}{Simon W{\"o}rner},
  {and} \bibinfo{person}{Thorsten Holz}.} \bibinfo{year}{2020}\natexlab{}.
\newblock \showarticletitle{{AURORA}: Statistical Crash Analysis for Automated
  Root Cause Explanation}. In \bibinfo{booktitle}{\emph{29th {USENIX} Security
  Symposium ({USENIX} Security 20)}}. \bibinfo{publisher}{{USENIX}
  Association}, \bibinfo{pages}{235--252}.
\newblock
\showISBNx{978-1-939133-17-5}
\urldef\tempurl%
\url{https://www.usenix.org/conference/usenixsecurity20/presentation/blazytko}
\showURL{%
\tempurl}


\bibitem[\protect\citeauthoryear{B{\"o}hme and Falk}{B{\"o}hme and
  Falk}{2020}]%
        {expcostbugs}
\bibfield{author}{\bibinfo{person}{Marcel B{\"o}hme} {and}
  \bibinfo{person}{Brandon Falk}.} \bibinfo{year}{2020}\natexlab{}.
\newblock \showarticletitle{Fuzzing: On the Exponential Cost of Vulnerability
  Discovery}. In \bibinfo{booktitle}{\emph{Proceedings of the 2020 28th ACM
  Joint Meeting on European Software Engineering Conference and Symposium on
  the Foundations of Software Engineering}} \emph{(\bibinfo{series}{ESEC/FSE
  2020})}. \bibinfo{publisher}{ACM}, \bibinfo{address}{New York, NY, USA}.
\newblock
\urldef\tempurl%
\url{https://doi.org/10.1145/3368089.3409729}
\showDOI{\tempurl}


\bibitem[\protect\citeauthoryear{B\"{o}hme, Pham, and Roychoudhury}{B\"{o}hme
  et~al\mbox{.}}{2016}]%
        {aflfast}
\bibfield{author}{\bibinfo{person}{Marcel B\"{o}hme},
  \bibinfo{person}{Van-Thuan Pham}, {and} \bibinfo{person}{Abhik
  Roychoudhury}.} \bibinfo{year}{2016}\natexlab{}.
\newblock \showarticletitle{Coverage-based Greybox Fuzzing As Markov Chain}. In
  \bibinfo{booktitle}{\emph{Proceedings of the 2016 ACM SIGSAC Conference on
  Computer and Communications Security}} (Vienna, Austria)
  \emph{(\bibinfo{series}{CCS '16})}. \bibinfo{publisher}{ACM},
  \bibinfo{address}{New York, NY, USA}, \bibinfo{pages}{1032--1043}.
\newblock
\showISBNx{978-1-4503-4139-4}
\urldef\tempurl%
\url{https://doi.org/10.1145/2976749.2978428}
\showDOI{\tempurl}


\bibitem[\protect\citeauthoryear{Caswell}{Caswell}{[n.d.]}]%
        {cgc}
\bibfield{author}{\bibinfo{person}{Brian Caswell}.}
  \bibinfo{year}{[n.d.]}\natexlab{}.
\newblock \bibinfo{title}{{C}yber {G}rand {C}hallenge Corpus}.
\newblock \bibinfo{howpublished}{\url{http://www.lungetech.com/cgc-corpus/}}.
\newblock


\bibitem[\protect\citeauthoryear{Chen and Chen}{Chen and Chen}{2018}]%
        {angora}
\bibfield{author}{\bibinfo{person}{P. Chen} {and} \bibinfo{person}{H. Chen}.}
  \bibinfo{year}{2018}\natexlab{}.
\newblock \showarticletitle{Angora: Efficient Fuzzing by Principled Search}. In
  \bibinfo{booktitle}{\emph{2018 IEEE Symposium on Security and Privacy (SP)}}.
  \bibinfo{publisher}{IEEE Computer Society}, \bibinfo{address}{Los Alamitos,
  CA, USA}, \bibinfo{pages}{711--725}.
\newblock
\showISSN{2375-1207}
\urldef\tempurl%
\url{https://doi.org/10.1109/SP.2018.00046}
\showDOI{\tempurl}


\bibitem[\protect\citeauthoryear{{Coppik}, {Schwahn}, and {Suri}}{{Coppik}
  et~al\mbox{.}}{2019}]%
        {memfuzz}
\bibfield{author}{\bibinfo{person}{N. {Coppik}}, \bibinfo{person}{O.
  {Schwahn}}, {and} \bibinfo{person}{N. {Suri}}.}
  \bibinfo{year}{2019}\natexlab{}.
\newblock \showarticletitle{MemFuzz: Using Memory Accesses to Guide Fuzzing}.
  In \bibinfo{booktitle}{\emph{2019 12th IEEE Conference on Software Testing,
  Validation and Verification (ICST)}}. \bibinfo{pages}{48--58}.
\newblock
\showISSN{2159-4848}
\urldef\tempurl%
\url{https://doi.org/10.1109/ICST.2019.00015}
\showDOI{\tempurl}


\bibitem[\protect\citeauthoryear{{Dolan-Gavitt}, {Hulin}, {Kirda}, {Leek},
  {Mambretti}, {Robertson}, {Ulrich}, and {Whelan}}{{Dolan-Gavitt}
  et~al\mbox{.}}{2016}]%
        {lavam}
\bibfield{author}{\bibinfo{person}{B. {Dolan-Gavitt}}, \bibinfo{person}{P.
  {Hulin}}, \bibinfo{person}{E. {Kirda}}, \bibinfo{person}{T. {Leek}},
  \bibinfo{person}{A. {Mambretti}}, \bibinfo{person}{W. {Robertson}},
  \bibinfo{person}{F. {Ulrich}}, {and} \bibinfo{person}{R. {Whelan}}.}
  \bibinfo{year}{2016}\natexlab{}.
\newblock \showarticletitle{LAVA: Large-Scale Automated Vulnerability
  Addition}. In \bibinfo{booktitle}{\emph{2016 IEEE Symposium on Security and
  Privacy (SP)}}. \bibinfo{pages}{110--121}.
\newblock
\showISSN{2375-1207}
\urldef\tempurl%
\url{https://doi.org/10.1109/SP.2016.15}
\showDOI{\tempurl}


\bibitem[\protect\citeauthoryear{{Dullien}}{{Dullien}}{2020}]%
        {weirdmachines}
\bibfield{author}{\bibinfo{person}{T. {Dullien}}.}
  \bibinfo{year}{2020}\natexlab{}.
\newblock \showarticletitle{Weird Machines, Exploitability, and Provable
  Unexploitability}.
\newblock \bibinfo{journal}{\emph{IEEE Transactions on Emerging Topics in
  Computing}} \bibinfo{volume}{8}, \bibinfo{number}{2} (\bibinfo{year}{2020}),
  \bibinfo{pages}{391--403}.
\newblock


\bibitem[\protect\citeauthoryear{Fioraldi, Maier, Ei{\ss}feldt, and
  Heuse}{Fioraldi et~al\mbox{.}}{2020}]%
        {aflplusplus}
\bibfield{author}{\bibinfo{person}{Andrea Fioraldi}, \bibinfo{person}{Dominik
  Maier}, \bibinfo{person}{Heiko Ei{\ss}feldt}, {and} \bibinfo{person}{Marc
  Heuse}.} \bibinfo{year}{2020}\natexlab{}.
\newblock \showarticletitle{AFL++ : Combining Incremental Steps of Fuzzing
  Research}. In \bibinfo{booktitle}{\emph{14th {USENIX} Workshop on Offensive
  Technologies ({WOOT} 20)}}. \bibinfo{publisher}{{USENIX} Association}.
\newblock
\urldef\tempurl%
\url{https://www.usenix.org/conference/woot20/presentation/fioraldi}
\showURL{%
\tempurl}
\newblock
\shownote{{Accessed: 2020-10-19}.}


\bibitem[\protect\citeauthoryear{Ganesh, Leek, and Rinard}{Ganesh
  et~al\mbox{.}}{2009}]%
        {buzzfuzz}
\bibfield{author}{\bibinfo{person}{Vijay Ganesh}, \bibinfo{person}{Tim Leek},
  {and} \bibinfo{person}{Martin~C. Rinard}.} \bibinfo{year}{2009}\natexlab{}.
\newblock \showarticletitle{Taint-based directed whitebox fuzzing}. In
  \bibinfo{booktitle}{\emph{31st International Conference on Software
  Engineering, {ICSE} 2009, May 16-24, 2009, Vancouver, Canada, Proceedings}}.
  \bibinfo{publisher}{{IEEE}}, \bibinfo{pages}{474--484}.
\newblock
\urldef\tempurl%
\url{https://doi.org/10.1109/ICSE.2009.5070546}
\showDOI{\tempurl}


\bibitem[\protect\citeauthoryear{Godefroid, Kiezun, and Levin}{Godefroid
  et~al\mbox{.}}{2008}]%
        {gode2008}
\bibfield{author}{\bibinfo{person}{Patrice Godefroid}, \bibinfo{person}{Adam
  Kiezun}, {and} \bibinfo{person}{Michael~Y. Levin}.}
  \bibinfo{year}{2008}\natexlab{}.
\newblock \showarticletitle{Grammar-based whitebox fuzzing}. In
  \bibinfo{booktitle}{\emph{Proceedings of the {ACM} {SIGPLAN} 2008 Conference
  on Programming Language Design and Implementation, Tucson, AZ, USA, June
  7-13, 2008}}, \bibfield{editor}{\bibinfo{person}{Rajiv Gupta} {and}
  \bibinfo{person}{Saman~P. Amarasinghe}} (Eds.). \bibinfo{publisher}{{ACM}},
  \bibinfo{pages}{206--215}.
\newblock
\urldef\tempurl%
\url{https://doi.org/10.1145/1375581.1375607}
\showDOI{\tempurl}


\bibitem[\protect\citeauthoryear{{Google}}{{Google}}{[n.d.]a}]%
        {fuzzbench}
\bibfield{author}{\bibinfo{person}{{Google}}.}
  \bibinfo{year}{[n.d.]}\natexlab{a}.
\newblock \bibinfo{title}{{FuzzBench}}.
\newblock \bibinfo{howpublished}{\url{https://google.github.io/fuzzbench/}}.
\newblock
\newblock
\shownote{{Accessed: 2020-05-02}.}


\bibitem[\protect\citeauthoryear{{Google}}{{Google}}{[n.d.]b}]%
        {gfuzzsuite}
\bibfield{author}{\bibinfo{person}{{Google}}.}
  \bibinfo{year}{[n.d.]}\natexlab{b}.
\newblock \bibinfo{title}{{Fuzzer Test Suite}}.
\newblock
  \bibinfo{howpublished}{\url{https://github.com/google/fuzzer-test-suite}}.
\newblock
\newblock
\shownote{{Accessed: 2019-09-06}.}


\bibitem[\protect\citeauthoryear{Google}{Google}{[n.d.]}]%
        {honggfuzz}
\bibfield{author}{\bibinfo{person}{Google}.} \bibinfo{year}{[n.d.]}\natexlab{}.
\newblock \bibinfo{title}{{honggfuzz}}.
\newblock \bibinfo{howpublished}{\url{http://honggfuzz.com/}}.
\newblock
\newblock
\shownote{{Accessed: 2019-10-19}.}


\bibitem[\protect\citeauthoryear{Grieco, Ceresa, and Buiras}{Grieco
  et~al\mbox{.}}{2016}]%
        {quickfuzz}
\bibfield{author}{\bibinfo{person}{Gustavo Grieco},
  \bibinfo{person}{Mart{\'{\i}}n Ceresa}, {and} \bibinfo{person}{Pablo
  Buiras}.} \bibinfo{year}{2016}\natexlab{}.
\newblock \showarticletitle{QuickFuzz: an automatic random fuzzer for common
  file formats}. In \bibinfo{booktitle}{\emph{Proceedings of the 9th
  International Symposium on Haskell, Haskell 2016, Nara, Japan, September
  22-23, 2016}}, \bibfield{editor}{\bibinfo{person}{Geoffrey Mainland}} (Ed.).
  \bibinfo{publisher}{{ACM}}, \bibinfo{pages}{13--20}.
\newblock
\urldef\tempurl%
\url{https://doi.org/10.1145/2976002.2976017}
\showDOI{\tempurl}


\bibitem[\protect\citeauthoryear{Gulwani}{Gulwani}{2010}]%
        {progsynth}
\bibfield{author}{\bibinfo{person}{Sumit Gulwani}.}
  \bibinfo{year}{2010}\natexlab{}.
\newblock \showarticletitle{Dimensions in Program Synthesis}. In
  \bibinfo{booktitle}{\emph{Proceedings of the 12th International ACM SIGPLAN
  Symposium on Principles and Practice of Declarative Programming}} (Hagenberg,
  Austria) \emph{(\bibinfo{series}{PPDP ’10})}.
  \bibinfo{publisher}{Association for Computing Machinery},
  \bibinfo{address}{New York, NY, USA}, \bibinfo{pages}{13–24}.
\newblock
\showISBNx{9781450301329}
\urldef\tempurl%
\url{https://doi.org/10.1145/1836089.1836091}
\showDOI{\tempurl}


\bibitem[\protect\citeauthoryear{Henning}{Henning}{2000}]%
        {speccpu2000}
\bibfield{author}{\bibinfo{person}{John~L. Henning}.}
  \bibinfo{year}{2000}\natexlab{}.
\newblock \showarticletitle{SPEC CPU2000: Measuring CPU Performance in the New
  Millennium}.
\newblock \bibinfo{journal}{\emph{Computer}} \bibinfo{volume}{33},
  \bibinfo{number}{7} (\bibinfo{date}{July} \bibinfo{year}{2000}),
  \bibinfo{pages}{28–35}.
\newblock
\showISSN{0018-9162}
\urldef\tempurl%
\url{https://doi.org/10.1109/2.869367}
\showDOI{\tempurl}


\bibitem[\protect\citeauthoryear{Intel}{Intel}{[n.d.]}]%
        {intelpinapi}
\bibfield{author}{\bibinfo{person}{Intel}.} \bibinfo{year}{[n.d.]}\natexlab{}.
\newblock \bibinfo{title}{Intel Pin API Reference}.
\newblock
  \bibinfo{howpublished}{\url{https://software.intel.com/sites/landingpage/pintool/docs/71313/Pin/html/index.html}}.
\newblock


\bibitem[\protect\citeauthoryear{Ispoglou}{Ispoglou}{2020}]%
        {fuzzgen}
\bibfield{author}{\bibinfo{person}{Kyriakos~K. Ispoglou}.}
  \bibinfo{year}{2020}\natexlab{}.
\newblock \showarticletitle{FuzzGen: Automatic Fuzzer Generation}. In
  \bibinfo{booktitle}{\emph{Proceedings of the USENIX Conference on Security
  Symposium}}.
\newblock


\bibitem[\protect\citeauthoryear{Joshi, Phansalkar, Eeckhout, and John}{Joshi
  et~al\mbox{.}}{2006}]%
        {benchmarksimilarity}
\bibfield{author}{\bibinfo{person}{Ajay Joshi}, \bibinfo{person}{Aashish
  Phansalkar}, \bibinfo{person}{L. Eeckhout}, {and} \bibinfo{person}{L.~K.
  John}.} \bibinfo{year}{2006}\natexlab{}.
\newblock \showarticletitle{Measuring benchmark similarity using inherent
  program characteristics}.
\newblock \bibinfo{journal}{\emph{IEEE Trans. Comput.}} \bibinfo{volume}{55},
  \bibinfo{number}{6} (\bibinfo{year}{2006}), \bibinfo{pages}{769--782}.
\newblock


\bibitem[\protect\citeauthoryear{Kaplan and Meier}{Kaplan and Meier}{1958}]%
        {kaplanmeier}
\bibfield{author}{\bibinfo{person}{Edward~L Kaplan} {and} \bibinfo{person}{Paul
  Meier}.} \bibinfo{year}{1958}\natexlab{}.
\newblock \showarticletitle{Nonparametric estimation from incomplete
  observations}.
\newblock \bibinfo{journal}{\emph{Journal of the American statistical
  association}} \bibinfo{volume}{53}, \bibinfo{number}{282}
  (\bibinfo{year}{1958}), \bibinfo{pages}{457--481}.
\newblock


\bibitem[\protect\citeauthoryear{{Kashyap}, {Ruchti}, {Kot}, {Turetsky},
  {Swords}, {Pan}, {Henry}, {Melski}, and {Schulte}}{{Kashyap}
  et~al\mbox{.}}{2019}]%
        {buginjector}
\bibfield{author}{\bibinfo{person}{V. {Kashyap}}, \bibinfo{person}{J.
  {Ruchti}}, \bibinfo{person}{L. {Kot}}, \bibinfo{person}{E. {Turetsky}},
  \bibinfo{person}{R. {Swords}}, \bibinfo{person}{S.~A. {Pan}},
  \bibinfo{person}{J. {Henry}}, \bibinfo{person}{D. {Melski}}, {and}
  \bibinfo{person}{E. {Schulte}}.} \bibinfo{year}{2019}\natexlab{}.
\newblock \showarticletitle{Automated Customized Bug-Benchmark Generation}. In
  \bibinfo{booktitle}{\emph{2019 19th International Working Conference on
  Source Code Analysis and Manipulation (SCAM)}}. \bibinfo{pages}{103--114}.
\newblock


\bibitem[\protect\citeauthoryear{Klees, Ruef, Cooper, Wei, and Hicks}{Klees
  et~al\mbox{.}}{2018}]%
        {fuzzeval}
\bibfield{author}{\bibinfo{person}{George Klees}, \bibinfo{person}{Andrew
  Ruef}, \bibinfo{person}{Benji Cooper}, \bibinfo{person}{Shiyi Wei}, {and}
  \bibinfo{person}{Michael Hicks}.} \bibinfo{year}{2018}\natexlab{}.
\newblock \showarticletitle{Evaluating Fuzz Testing}. In
  \bibinfo{booktitle}{\emph{Proceedings of the 2018 ACM SIGSAC Conference on
  Computer and Communications Security}} (Toronto, Canada)
  \emph{(\bibinfo{series}{CCS '18})}. \bibinfo{publisher}{ACM},
  \bibinfo{address}{New York, NY, USA}, \bibinfo{pages}{2123--2138}.
\newblock
\showISBNx{978-1-4503-5693-0}
\urldef\tempurl%
\url{https://doi.org/10.1145/3243734.3243804}
\showDOI{\tempurl}


\bibitem[\protect\citeauthoryear{Lemieux and Sen}{Lemieux and Sen}{2018}]%
        {fairfuzz}
\bibfield{author}{\bibinfo{person}{Caroline Lemieux} {and}
  \bibinfo{person}{Koushik Sen}.} \bibinfo{year}{2018}\natexlab{}.
\newblock \showarticletitle{FairFuzz: a targeted mutation strategy for
  increasing greybox fuzz testing coverage}. In
  \bibinfo{booktitle}{\emph{Proceedings of the 33rd {ACM/IEEE} International
  Conference on Automated Software Engineering, {ASE} 2018, Montpellier,
  France, September 3-7, 2018}}, \bibfield{editor}{\bibinfo{person}{Marianne
  Huchard}, \bibinfo{person}{Christian K{\"{a}}stner}, {and}
  \bibinfo{person}{Gordon Fraser}} (Eds.). \bibinfo{publisher}{{ACM}},
  \bibinfo{pages}{475--485}.
\newblock
\urldef\tempurl%
\url{https://doi.org/10.1145/3238147.3238176}
\showDOI{\tempurl}


\bibitem[\protect\citeauthoryear{Li, Chen, Chandramohan, Lin, Liu, and Tiu}{Li
  et~al\mbox{.}}{2017}]%
        {steelix}
\bibfield{author}{\bibinfo{person}{Yuekang Li}, \bibinfo{person}{Bihuan Chen},
  \bibinfo{person}{Mahinthan Chandramohan}, \bibinfo{person}{Shang-Wei Lin},
  \bibinfo{person}{Yang Liu}, {and} \bibinfo{person}{Alwen Tiu}.}
  \bibinfo{year}{2017}\natexlab{}.
\newblock \showarticletitle{Steelix: Program-State Based Binary Fuzzing}. In
  \bibinfo{booktitle}{\emph{Proceedings of the 2017 11th Joint Meeting on
  Foundations of Software Engineering}} (Paderborn, Germany)
  \emph{(\bibinfo{series}{ESEC/FSE 2017})}. \bibinfo{publisher}{Association for
  Computing Machinery}, \bibinfo{address}{New York, NY, USA},
  \bibinfo{pages}{627–637}.
\newblock
\showISBNx{9781450351058}
\urldef\tempurl%
\url{https://doi.org/10.1145/3106237.3106295}
\showDOI{\tempurl}


\bibitem[\protect\citeauthoryear{Li, Ji, Chen, Liang, Lee, Chen, Lyu, Wu,
  Beyah, Cheng, Lu, and Wang}{Li et~al\mbox{.}}{2021}]%
        {unifuzz}
\bibfield{author}{\bibinfo{person}{Yuwei Li}, \bibinfo{person}{Shouling Ji},
  \bibinfo{person}{Yuan Chen}, \bibinfo{person}{Sizhuang Liang},
  \bibinfo{person}{Wei-Han Lee}, \bibinfo{person}{Yueyao Chen},
  \bibinfo{person}{Chenyang Lyu}, \bibinfo{person}{Chunming Wu},
  \bibinfo{person}{Raheem Beyah}, \bibinfo{person}{Peng Cheng},
  \bibinfo{person}{Kangjie Lu}, {and} \bibinfo{person}{Ting Wang}.}
  \bibinfo{year}{2021}\natexlab{}.
\newblock \showarticletitle{UNIFUZZ: A Holistic and Pragmatic Metrics-Driven
  Platform for Evaluating Fuzzers}. In \bibinfo{booktitle}{\emph{30th {USENIX}
  Security Symposium ({USENIX} Security 21)}}. \bibinfo{publisher}{{USENIX}
  Association}.
\newblock


\bibitem[\protect\citeauthoryear{{LLVM Foundation}}{{LLVM
  Foundation}}{[n.d.]}]%
        {libfuzzer}
\bibfield{author}{\bibinfo{person}{{LLVM Foundation}}.}
  \bibinfo{year}{[n.d.]}\natexlab{}.
\newblock \bibinfo{title}{{libFuzzer}}.
\newblock \bibinfo{howpublished}{\url{https://llvm.org/docs/LibFuzzer.html}}.
\newblock
\newblock
\shownote{{Accessed: 2019-09-06}.}


\bibitem[\protect\citeauthoryear{Lu, Li, Qin, Tan, Zhou, and Zhou}{Lu
  et~al\mbox{.}}{2005}]%
        {bugbench}
\bibfield{author}{\bibinfo{person}{Shan Lu}, \bibinfo{person}{Zhenmin Li},
  \bibinfo{person}{Feng Qin}, \bibinfo{person}{Lin Tan}, \bibinfo{person}{Pin
  Zhou}, {and} \bibinfo{person}{Yuanyuan Zhou}.}
  \bibinfo{year}{2005}\natexlab{}.
\newblock \showarticletitle{Bugbench: Benchmarks for evaluating bug detection
  tools}. In \bibinfo{booktitle}{\emph{In Workshop on the Evaluation of
  Software Defect Detection Tools}}.
\newblock


\bibitem[\protect\citeauthoryear{Luk, Cohn, Muth, Patil, Klauser, Lowney,
  Wallace, Reddi, and Hazelwood}{Luk et~al\mbox{.}}{2005}]%
        {intelpin}
\bibfield{author}{\bibinfo{person}{Chi-Keung Luk}, \bibinfo{person}{Robert
  Cohn}, \bibinfo{person}{Robert Muth}, \bibinfo{person}{Harish Patil},
  \bibinfo{person}{Artur Klauser}, \bibinfo{person}{Geoff Lowney},
  \bibinfo{person}{Steven Wallace}, \bibinfo{person}{Vijay~Janapa Reddi}, {and}
  \bibinfo{person}{Kim Hazelwood}.} \bibinfo{year}{2005}\natexlab{}.
\newblock \showarticletitle{Pin: Building Customized Program Analysis Tools
  with Dynamic Instrumentation}. In \bibinfo{booktitle}{\emph{Proceedings of
  the 2005 ACM SIGPLAN Conference on Programming Language Design and
  Implementation}} (Chicago, IL, USA) \emph{(\bibinfo{series}{PLDI '05})}.
  \bibinfo{publisher}{Association for Computing Machinery},
  \bibinfo{address}{New York, NY, USA}, \bibinfo{pages}{190–200}.
\newblock
\showISBNx{1595930566}
\urldef\tempurl%
\url{https://doi.org/10.1145/1065010.1065034}
\showDOI{\tempurl}


\bibitem[\protect\citeauthoryear{Lyu, Ji, Zhang, Li, Lee, Song, and Beyah}{Lyu
  et~al\mbox{.}}{2019}]%
        {moptafl}
\bibfield{author}{\bibinfo{person}{Chenyang Lyu}, \bibinfo{person}{Shouling
  Ji}, \bibinfo{person}{Chao Zhang}, \bibinfo{person}{Yuwei Li},
  \bibinfo{person}{Wei{-}Han Lee}, \bibinfo{person}{Yu Song}, {and}
  \bibinfo{person}{Raheem Beyah}.} \bibinfo{year}{2019}\natexlab{}.
\newblock \showarticletitle{{MOPT:} Optimized Mutation Scheduling for Fuzzers}.
  In \bibinfo{booktitle}{\emph{28th {USENIX} Security Symposium, {USENIX}
  Security 2019, Santa Clara, CA, USA, August 14-16, 2019.}},
  \bibfield{editor}{\bibinfo{person}{Nadia Heninger} {and}
  \bibinfo{person}{Patrick Traynor}} (Eds.). \bibinfo{publisher}{{USENIX}
  Association}, \bibinfo{pages}{1949--1966}.
\newblock
\urldef\tempurl%
\url{https://www.usenix.org/conference/usenixsecurity19/presentation/lyu}
\showURL{%
\tempurl}


\bibitem[\protect\citeauthoryear{Man{\`{e}}s, Han, Han, Cha, Egele, Schwartz,
  and Woo}{Man{\`{e}}s et~al\mbox{.}}{2019}]%
        {fuzzingart}
\bibfield{author}{\bibinfo{person}{Valentin J.~M. Man{\`{e}}s},
  \bibinfo{person}{HyungSeok Han}, \bibinfo{person}{Choongwoo Han},
  \bibinfo{person}{Sang~Kil Cha}, \bibinfo{person}{Manuel Egele},
  \bibinfo{person}{Edward~J. Schwartz}, {and} \bibinfo{person}{Maverick Woo}.}
  \bibinfo{year}{2019}\natexlab{}.
\newblock \showarticletitle{The Art, Science, and Engineering of Fuzzing: A
  Survey}.
\newblock \bibinfo{journal}{\emph{IEEE Transactions on Software Engineering}}
  (\bibinfo{year}{2019}).
\newblock
\urldef\tempurl%
\url{https://doi.org/10.1109/TSE.2019.2946563}
\showDOI{\tempurl}


\bibitem[\protect\citeauthoryear{Man\`{e}s, Kim, and Cha}{Man\`{e}s
  et~al\mbox{.}}{2020}]%
        {ankou}
\bibfield{author}{\bibinfo{person}{Valentin J.~M. Man\`{e}s},
  \bibinfo{person}{Soomin Kim}, {and} \bibinfo{person}{Sang~Kil Cha}.}
  \bibinfo{year}{2020}\natexlab{}.
\newblock \showarticletitle{Ankou: Guiding Grey-Box Fuzzing towards
  Combinatorial Difference}. In \bibinfo{booktitle}{\emph{Proceedings of the
  ACM/IEEE 42nd International Conference on Software Engineering}} (Seoul,
  South Korea) \emph{(\bibinfo{series}{ICSE '20})}.
  \bibinfo{publisher}{Association for Computing Machinery},
  \bibinfo{address}{New York, NY, USA}, \bibinfo{pages}{1024–1036}.
\newblock
\showISBNx{9781450371216}
\urldef\tempurl%
\url{https://doi.org/10.1145/3377811.3380421}
\showDOI{\tempurl}


\bibitem[\protect\citeauthoryear{{MITRE}}{{MITRE}}{2007}]%
        {cwe}
\bibfield{author}{\bibinfo{person}{{MITRE}}.} \bibinfo{year}{2007}\natexlab{}.
\newblock \bibinfo{title}{Common Weakness Enumeration (CWE)}.
\newblock \bibinfo{howpublished}{\url{https://cwe.mitre.org/}}.
\newblock


\bibitem[\protect\citeauthoryear{Nosco, Ziegler, Clark, Marrero, Finkler,
  Barbarello, and Petullo}{Nosco et~al\mbox{.}}{2020}]%
        {industrialhacking}
\bibfield{author}{\bibinfo{person}{Timothy Nosco}, \bibinfo{person}{Jared
  Ziegler}, \bibinfo{person}{Zechariah Clark}, \bibinfo{person}{Davy Marrero},
  \bibinfo{person}{Todd Finkler}, \bibinfo{person}{Andrew Barbarello}, {and}
  \bibinfo{person}{W.~Michael Petullo}.} \bibinfo{year}{2020}\natexlab{}.
\newblock \showarticletitle{The Industrial Age of Hacking}. In
  \bibinfo{booktitle}{\emph{29th {USENIX} Security Symposium ({USENIX} Security
  20)}}. \bibinfo{publisher}{{USENIX} Association},
  \bibinfo{pages}{1129--1146}.
\newblock
\showISBNx{978-1-939133-17-5}
\urldef\tempurl%
\url{https://www.usenix.org/conference/usenixsecurity20/presentation/nosco}
\showURL{%
\tempurl}


\bibitem[\protect\citeauthoryear{{Peach Tech}}{{Peach Tech}}{[n.d.]}]%
        {peachfuzz}
\bibfield{author}{\bibinfo{person}{{Peach Tech}}.}
  \bibinfo{year}{[n.d.]}\natexlab{}.
\newblock \bibinfo{title}{Peach Fuzzer Platform}.
\newblock
  \bibinfo{howpublished}{\url{https://www.peach.tech/products/peach-fuzzer/peach-platform/}}.
\newblock
\newblock
\shownote{{Accessed: 2019-09-09}.}


\bibitem[\protect\citeauthoryear{Pearson}{Pearson}{1901}]%
        {pca}
\bibfield{author}{\bibinfo{person}{Karl Pearson}.}
  \bibinfo{year}{1901}\natexlab{}.
\newblock \showarticletitle{LIII. On lines and planes of closest fit to systems
  of points in space}.
\newblock \bibinfo{journal}{\emph{The London, Edinburgh, and Dublin
  Philosophical Magazine and Journal of Science}} \bibinfo{volume}{2},
  \bibinfo{number}{11} (\bibinfo{year}{1901}), \bibinfo{pages}{559--572}.
\newblock


\bibitem[\protect\citeauthoryear{{Peng}, {Shoshitaishvili}, and {Payer}}{{Peng}
  et~al\mbox{.}}{2018}]%
        {tfuzz}
\bibfield{author}{\bibinfo{person}{H. {Peng}}, \bibinfo{person}{Y.
  {Shoshitaishvili}}, {and} \bibinfo{person}{M. {Payer}}.}
  \bibinfo{year}{2018}\natexlab{}.
\newblock \showarticletitle{T-Fuzz: Fuzzing by Program Transformation}. In
  \bibinfo{booktitle}{\emph{2018 IEEE Symposium on Security and Privacy (SP)}}.
  \bibinfo{pages}{697--710}.
\newblock
\showISSN{2375-1207}
\urldef\tempurl%
\url{https://doi.org/10.1109/SP.2018.00056}
\showDOI{\tempurl}


\bibitem[\protect\citeauthoryear{{Petsios}, {Tang}, {Stolfo}, {Keromytis}, and
  {Jana}}{{Petsios} et~al\mbox{.}}{2017}]%
        {nezha}
\bibfield{author}{\bibinfo{person}{T. {Petsios}}, \bibinfo{person}{A. {Tang}},
  \bibinfo{person}{S. {Stolfo}}, \bibinfo{person}{A.~D. {Keromytis}}, {and}
  \bibinfo{person}{S. {Jana}}.} \bibinfo{year}{2017}\natexlab{}.
\newblock \showarticletitle{NEZHA: Efficient Domain-Independent Differential
  Testing}. In \bibinfo{booktitle}{\emph{2017 IEEE Symposium on Security and
  Privacy (SP)}}. \bibinfo{pages}{615--632}.
\newblock
\showISSN{2375-1207}
\urldef\tempurl%
\url{https://doi.org/10.1109/SP.2017.27}
\showDOI{\tempurl}


\bibitem[\protect\citeauthoryear{Petsios, Zhao, Keromytis, and Jana}{Petsios
  et~al\mbox{.}}{2017}]%
        {slowfuzz}
\bibfield{author}{\bibinfo{person}{Theofilos Petsios}, \bibinfo{person}{Jason
  Zhao}, \bibinfo{person}{Angelos~D. Keromytis}, {and} \bibinfo{person}{Suman
  Jana}.} \bibinfo{year}{2017}\natexlab{}.
\newblock \showarticletitle{SlowFuzz: Automated Domain-Independent Detection of
  Algorithmic Complexity Vulnerabilities}. In
  \bibinfo{booktitle}{\emph{Proceedings of the 2017 ACM SIGSAC Conference on
  Computer and Communications Security}} (Dallas, Texas, USA)
  \emph{(\bibinfo{series}{CCS '17})}. \bibinfo{publisher}{ACM},
  \bibinfo{address}{New York, NY, USA}, \bibinfo{pages}{2155--2168}.
\newblock
\showISBNx{978-1-4503-4946-8}
\urldef\tempurl%
\url{https://doi.org/10.1145/3133956.3134073}
\showDOI{\tempurl}


\bibitem[\protect\citeauthoryear{Pham, B{\"{o}}hme, and Roychoudhury}{Pham
  et~al\mbox{.}}{2016}]%
        {mowf}
\bibfield{author}{\bibinfo{person}{Van{-}Thuan Pham}, \bibinfo{person}{Marcel
  B{\"{o}}hme}, {and} \bibinfo{person}{Abhik Roychoudhury}.}
  \bibinfo{year}{2016}\natexlab{}.
\newblock \showarticletitle{Model-based whitebox fuzzing for program binaries}.
  In \bibinfo{booktitle}{\emph{Proceedings of the 31st {IEEE/ACM} International
  Conference on Automated Software Engineering, {ASE} 2016, Singapore,
  September 3-7, 2016}}, \bibfield{editor}{\bibinfo{person}{David Lo},
  \bibinfo{person}{Sven Apel}, {and} \bibinfo{person}{Sarfraz Khurshid}}
  (Eds.). \bibinfo{publisher}{{ACM}}, \bibinfo{pages}{543--553}.
\newblock
\urldef\tempurl%
\url{https://doi.org/10.1145/2970276.2970316}
\showDOI{\tempurl}


\bibitem[\protect\citeauthoryear{{Phansalkar}, {Joshi}, {Eeckhout}, and
  {John}}{{Phansalkar} et~al\mbox{.}}{2005}]%
        {programsimilarity}
\bibfield{author}{\bibinfo{person}{A. {Phansalkar}}, \bibinfo{person}{A.
  {Joshi}}, \bibinfo{person}{L. {Eeckhout}}, {and} \bibinfo{person}{L.~K.
  {John}}.} \bibinfo{year}{2005}\natexlab{}.
\newblock \showarticletitle{Measuring Program Similarity: Experiments with SPEC
  CPU Benchmark Suites}. In \bibinfo{booktitle}{\emph{IEEE International
  Symposium on Performance Analysis of Systems and Software, 2005. ISPASS
  2005.}} \bibinfo{pages}{10--20}.
\newblock


\bibitem[\protect\citeauthoryear{Poeplau and Francillon}{Poeplau and
  Francillon}{2020}]%
        {symcc}
\bibfield{author}{\bibinfo{person}{Sebastian Poeplau} {and}
  \bibinfo{person}{Aur{\'e}lien Francillon}.} \bibinfo{year}{2020}\natexlab{}.
\newblock \showarticletitle{Symbolic execution with SymCC:
  Don{\textquoteright}t interpret, compile!}. In \bibinfo{booktitle}{\emph{29th
  {USENIX} Security Symposium ({USENIX} Security 20)}}.
  \bibinfo{publisher}{{USENIX} Association}, \bibinfo{pages}{181--198}.
\newblock
\showISBNx{978-1-939133-17-5}
\urldef\tempurl%
\url{https://www.usenix.org/conference/usenixsecurity20/presentation/poeplau}
\showURL{%
\tempurl}


\bibitem[\protect\citeauthoryear{Prokopec, Ros\`{a}, Leopoldseder, Duboscq,
  T\r{u}ma, Studener, Bulej, Zheng, Villaz\'{o}n, Simon, W\"{u}rthinger, and
  Binder}{Prokopec et~al\mbox{.}}{2019}]%
        {renaissance}
\bibfield{author}{\bibinfo{person}{Aleksandar Prokopec},
  \bibinfo{person}{Andrea Ros\`{a}}, \bibinfo{person}{David Leopoldseder},
  \bibinfo{person}{Gilles Duboscq}, \bibinfo{person}{Petr T\r{u}ma},
  \bibinfo{person}{Martin Studener}, \bibinfo{person}{Lubom\'{\i}r Bulej},
  \bibinfo{person}{Yudi Zheng}, \bibinfo{person}{Alex Villaz\'{o}n},
  \bibinfo{person}{Doug Simon}, \bibinfo{person}{Thomas W\"{u}rthinger}, {and}
  \bibinfo{person}{Walter Binder}.} \bibinfo{year}{2019}\natexlab{}.
\newblock \showarticletitle{Renaissance: Benchmarking Suite for Parallel
  Applications on the JVM}. In \bibinfo{booktitle}{\emph{Proceedings of the
  40th ACM SIGPLAN Conference on Programming Language Design and
  Implementation}} (Phoenix, AZ, USA) \emph{(\bibinfo{series}{PLDI 2019})}.
  \bibinfo{publisher}{Association for Computing Machinery},
  \bibinfo{address}{New York, NY, USA}, \bibinfo{pages}{31–47}.
\newblock
\showISBNx{9781450367127}
\urldef\tempurl%
\url{https://doi.org/10.1145/3314221.3314637}
\showDOI{\tempurl}


\bibitem[\protect\citeauthoryear{Rains}{Rains}{2012}]%
        {msftfuzz}
\bibfield{author}{\bibinfo{person}{Tim Rains}.}
  \bibinfo{year}{2012}\natexlab{}.
\newblock \bibinfo{title}{Security Development Lifecycle: A Living Process}.
\newblock
  \bibinfo{howpublished}{\url{https://www.microsoft.com/security/blog/2012/02/01/security-development-lifecycle-a-living-process/}}.
\newblock
\newblock
\shownote{{Accessed: 2019-09-09}.}


\bibitem[\protect\citeauthoryear{Roy, Pandey, Dolan-Gavitt, and Hu}{Roy
  et~al\mbox{.}}{2018}]%
        {bug-synthesis}
\bibfield{author}{\bibinfo{person}{Subhajit Roy}, \bibinfo{person}{Awanish
  Pandey}, \bibinfo{person}{Brendan Dolan-Gavitt}, {and} \bibinfo{person}{Yu
  Hu}.} \bibinfo{year}{2018}\natexlab{}.
\newblock \showarticletitle{Bug Synthesis: Challenging Bug-Finding Tools with
  Deep Faults}. In \bibinfo{booktitle}{\emph{Proceedings of the 2018 26th ACM
  Joint Meeting on European Software Engineering Conference and Symposium on
  the Foundations of Software Engineering}} (Lake Buena Vista, FL, USA)
  \emph{(\bibinfo{series}{ESEC/FSE 2018})}. \bibinfo{publisher}{Association for
  Computing Machinery}, \bibinfo{address}{New York, NY, USA},
  \bibinfo{pages}{224–234}.
\newblock
\showISBNx{9781450355735}
\urldef\tempurl%
\url{https://doi.org/10.1145/3236024.3236084}
\showDOI{\tempurl}


\bibitem[\protect\citeauthoryear{Serebryany, Bruening, Potapenko, and
  Vyukov}{Serebryany et~al\mbox{.}}{2012}]%
        {asan}
\bibfield{author}{\bibinfo{person}{Konstantin Serebryany},
  \bibinfo{person}{Derek Bruening}, \bibinfo{person}{Alexander Potapenko},
  {and} \bibinfo{person}{Dmitriy Vyukov}.} \bibinfo{year}{2012}\natexlab{}.
\newblock \showarticletitle{AddressSanitizer: {A} Fast Address Sanity Checker}.
  In \bibinfo{booktitle}{\emph{2012 {USENIX} Annual Technical Conference,
  Boston, MA, USA, June 13-15, 2012}},
  \bibfield{editor}{\bibinfo{person}{Gernot Heiser} {and}
  \bibinfo{person}{Wilson~C. Hsieh}} (Eds.). \bibinfo{publisher}{{USENIX}
  Association}, \bibinfo{pages}{309--318}.
\newblock
\urldef\tempurl%
\url{https://www.usenix.org/conference/atc12/technical-sessions/presentation/serebryany}
\showURL{%
\tempurl}


\bibitem[\protect\citeauthoryear{Shoshitaishvili, Wang, Salls, Stephens,
  Polino, Dutcher, Grosen, Feng, Hauser, Kruegel, and Vigna}{Shoshitaishvili
  et~al\mbox{.}}{2016}]%
        {angr}
\bibfield{author}{\bibinfo{person}{Yan Shoshitaishvili}, \bibinfo{person}{Ruoyu
  Wang}, \bibinfo{person}{Christopher Salls}, \bibinfo{person}{Nick Stephens},
  \bibinfo{person}{Mario Polino}, \bibinfo{person}{Audrey Dutcher},
  \bibinfo{person}{John Grosen}, \bibinfo{person}{Siji Feng},
  \bibinfo{person}{Christophe Hauser}, \bibinfo{person}{Christopher Kruegel},
  {and} \bibinfo{person}{Giovanni Vigna}.} \bibinfo{year}{2016}\natexlab{}.
\newblock \showarticletitle{{SoK: (State of) The Art of War: Offensive
  Techniques in Binary Analysis}}. In \bibinfo{booktitle}{\emph{IEEE Symposium
  on Security and Privacy}}.
\newblock


\bibitem[\protect\citeauthoryear{{Song}, {Lettner}, {Rajasekaran}, {Na},
  {Volckaert}, {Larsen}, and {Franz}}{{Song} et~al\mbox{.}}{2019}]%
        {sanitizers}
\bibfield{author}{\bibinfo{person}{D. {Song}}, \bibinfo{person}{J. {Lettner}},
  \bibinfo{person}{P. {Rajasekaran}}, \bibinfo{person}{Y. {Na}},
  \bibinfo{person}{S. {Volckaert}}, \bibinfo{person}{P. {Larsen}}, {and}
  \bibinfo{person}{M. {Franz}}.} \bibinfo{year}{2019}\natexlab{}.
\newblock \showarticletitle{SoK: Sanitizing for Security}. In
  \bibinfo{booktitle}{\emph{2019 IEEE Symposium on Security and Privacy (SP)}}.
  \bibinfo{pages}{1275--1295}.
\newblock
\showISSN{2375-1207}
\urldef\tempurl%
\url{https://doi.org/10.1109/SP.2019.00010}
\showDOI{\tempurl}


\bibitem[\protect\citeauthoryear{Sprenkels}{Sprenkels}{[n.d.]}]%
        {llvmsidechannel}
\bibfield{author}{\bibinfo{person}{Daan Sprenkels}.}
  \bibinfo{year}{[n.d.]}\natexlab{}.
\newblock \bibinfo{title}{{LLVM provides no side-channel resistance}}.
\newblock
  \bibinfo{howpublished}{\url{https://dsprenkels.com/cmov-conversion.html}}.
\newblock
\newblock
\shownote{{Accessed: 2020-02-13}.}


\bibitem[\protect\citeauthoryear{{Standard Performance Evaluation
  Corporation}}{{Standard Performance Evaluation Corporation}}{[n.d.]}]%
        {specbms}
\bibfield{author}{\bibinfo{person}{{Standard Performance Evaluation
  Corporation}}.} \bibinfo{year}{[n.d.]}\natexlab{}.
\newblock \bibinfo{title}{{SPEC Benchmark Suite}}.
\newblock \bibinfo{howpublished}{\url{https://www.spec.org/}}.
\newblock
\newblock
\shownote{{Accessed: 2020-02-12}.}


\bibitem[\protect\citeauthoryear{Stephens, Grosen, Salls, Dutcher, Wang,
  Corbetta, Shoshitaishvili, Kruegel, and Vigna}{Stephens
  et~al\mbox{.}}{2016}]%
        {driller}
\bibfield{author}{\bibinfo{person}{Nick Stephens}, \bibinfo{person}{John
  Grosen}, \bibinfo{person}{Christopher Salls}, \bibinfo{person}{Andrew
  Dutcher}, \bibinfo{person}{Ruoyu Wang}, \bibinfo{person}{Jacopo Corbetta},
  \bibinfo{person}{Yan Shoshitaishvili}, \bibinfo{person}{Christopher Kruegel},
  {and} \bibinfo{person}{Giovanni Vigna}.} \bibinfo{year}{2016}\natexlab{}.
\newblock \showarticletitle{Driller: Augmenting Fuzzing Through Selective
  Symbolic Execution}. In \bibinfo{booktitle}{\emph{23rd Annual Network and
  Distributed System Security Symposium, {NDSS} 2016, San Diego, California,
  USA, February 21-24, 2016}}. \bibinfo{publisher}{The Internet Society}.
\newblock
\urldef\tempurl%
\url{http://dx.doi.org/10.14722/ndss.2016.23368}
\showURL{%
\tempurl}


\bibitem[\protect\citeauthoryear{{Trail of Bits}}{{Trail of Bits}}{[n.d.]}]%
        {cb-multios}
\bibfield{author}{\bibinfo{person}{{Trail of Bits}}.}
  \bibinfo{year}{[n.d.]}\natexlab{}.
\newblock \bibinfo{title}{DARPA Challenge Binaries on Linux, OS X, and
  Windows}.
\newblock
  \bibinfo{howpublished}{\url{https://github.com/trailofbits/cb-multios/}}.
\newblock
\newblock
\shownote{{Accessed: 2020-10-04}.}


\bibitem[\protect\citeauthoryear{v.~Kistowski, Arnold, Huppler, Lange, Henning,
  and Cao}{v.~Kistowski et~al\mbox{.}}{2015}]%
        {howtobuildabenchmark}
\bibfield{author}{\bibinfo{person}{J\'{o}akim v. Kistowski},
  \bibinfo{person}{Jeremy~A. Arnold}, \bibinfo{person}{Karl Huppler},
  \bibinfo{person}{Klaus-Dieter Lange}, \bibinfo{person}{John~L. Henning},
  {and} \bibinfo{person}{Paul Cao}.} \bibinfo{year}{2015}\natexlab{}.
\newblock \showarticletitle{How to Build a Benchmark}. In
  \bibinfo{booktitle}{\emph{Proceedings of the 6th ACM/SPEC International
  Conference on Performance Engineering}} (Austin, Texas, USA)
  \emph{(\bibinfo{series}{ICPE ’15})}. \bibinfo{publisher}{Association for
  Computing Machinery}, \bibinfo{address}{New York, NY, USA},
  \bibinfo{pages}{333–336}.
\newblock
\showISBNx{9781450332484}
\urldef\tempurl%
\url{https://doi.org/10.1145/2668930.2688819}
\showDOI{\tempurl}


\bibitem[\protect\citeauthoryear{Wagner}{Wagner}{2017}]%
        {wagnerphd}
\bibfield{author}{\bibinfo{person}{Jonas~Benedict Wagner}.}
  \bibinfo{year}{2017}\natexlab{}.
\newblock \showarticletitle{Elastic Program Transformations Automatically
  Optimizing the Reliability/Performance Trade-off in Systems Software}.
\newblock  (\bibinfo{year}{2017}), \bibinfo{pages}{149}.
\newblock
\urldef\tempurl%
\url{https://doi.org/10.5075/epfl-thesis-7745}
\showDOI{\tempurl}


\bibitem[\protect\citeauthoryear{Wang, Chen, Wei, and Liu}{Wang
  et~al\mbox{.}}{2017}]%
        {skyfire}
\bibfield{author}{\bibinfo{person}{Junjie Wang}, \bibinfo{person}{Bihuan Chen},
  \bibinfo{person}{Lei Wei}, {and} \bibinfo{person}{Yang Liu}.}
  \bibinfo{year}{2017}\natexlab{}.
\newblock \showarticletitle{Skyfire: Data-Driven Seed Generation for Fuzzing}.
  In \bibinfo{booktitle}{\emph{2017 {IEEE} Symposium on Security and Privacy,
  {SP} 2017, San Jose, CA, USA, May 22-26, 2017}}. \bibinfo{publisher}{{IEEE}
  Computer Society}, \bibinfo{pages}{579--594}.
\newblock
\urldef\tempurl%
\url{https://doi.org/10.1109/SP.2017.23}
\showDOI{\tempurl}


\bibitem[\protect\citeauthoryear{Wang, Chen, Wei, and Liu}{Wang
  et~al\mbox{.}}{2019}]%
        {superion}
\bibfield{author}{\bibinfo{person}{Junjie Wang}, \bibinfo{person}{Bihuan Chen},
  \bibinfo{person}{Lei Wei}, {and} \bibinfo{person}{Yang Liu}.}
  \bibinfo{year}{2019}\natexlab{}.
\newblock \showarticletitle{Superion: grammar-aware greybox fuzzing}. In
  \bibinfo{booktitle}{\emph{Proceedings of the 41st International Conference on
  Software Engineering, {ICSE} 2019, Montreal, QC, Canada, May 25-31, 2019}},
  \bibfield{editor}{\bibinfo{person}{Joanne~M. Atlee}, \bibinfo{person}{Tevfik
  Bultan}, {and} \bibinfo{person}{Jon Whittle}} (Eds.).
  \bibinfo{publisher}{{IEEE} / {ACM}}, \bibinfo{pages}{724--735}.
\newblock
\urldef\tempurl%
\url{https://doi.org/10.1109/ICSE.2019.00081}
\showDOI{\tempurl}


\bibitem[\protect\citeauthoryear{Woo, Cha, Gottlieb, and Brumley}{Woo
  et~al\mbox{.}}{2013}]%
        {fuzzsim}
\bibfield{author}{\bibinfo{person}{Maverick Woo}, \bibinfo{person}{Sang~Kil
  Cha}, \bibinfo{person}{Samantha Gottlieb}, {and} \bibinfo{person}{David
  Brumley}.} \bibinfo{year}{2013}\natexlab{}.
\newblock \showarticletitle{Scheduling black-box mutational fuzzing}. In
  \bibinfo{booktitle}{\emph{2013 {ACM} {SIGSAC} Conference on Computer and
  Communications Security, CCS'13, Berlin, Germany, November 4-8, 2013}},
  \bibfield{editor}{\bibinfo{person}{Ahmad{-}Reza Sadeghi},
  \bibinfo{person}{Virgil~D. Gligor}, {and} \bibinfo{person}{Moti Yung}}
  (Eds.). \bibinfo{publisher}{{ACM}}, \bibinfo{pages}{511--522}.
\newblock
\urldef\tempurl%
\url{https://doi.org/10.1145/2508859.2516736}
\showDOI{\tempurl}


\bibitem[\protect\citeauthoryear{Yun, Lee, Xu, Jang, and Kim}{Yun
  et~al\mbox{.}}{2018}]%
        {qsym}
\bibfield{author}{\bibinfo{person}{Insu Yun}, \bibinfo{person}{Sangho Lee},
  \bibinfo{person}{Meng Xu}, \bibinfo{person}{Yeongjin Jang}, {and}
  \bibinfo{person}{Taesoo Kim}.} \bibinfo{year}{2018}\natexlab{}.
\newblock \showarticletitle{QSYM: A Practical Concolic Execution Engine
  Tailored for Hybrid Fuzzing}. In \bibinfo{booktitle}{\emph{Proceedings of the
  27th USENIX Conference on Security Symposium}} (Baltimore, MD, USA)
  \emph{(\bibinfo{series}{SEC'18})}. \bibinfo{publisher}{USENIX Association},
  \bibinfo{address}{Berkeley, CA, USA}, \bibinfo{pages}{745--761}.
\newblock
\showISBNx{978-1-931971-46-1}
\urldef\tempurl%
\url{http://dl.acm.org/citation.cfm?id=3277203.3277260}
\showURL{%
\tempurl}


\bibitem[\protect\citeauthoryear{Zalewski}{Zalewski}{[n.d.]}]%
        {afl}
\bibfield{author}{\bibinfo{person}{Michal Zalewski}.}
  \bibinfo{year}{[n.d.]}\natexlab{}.
\newblock \bibinfo{title}{{American Fuzzy Lop (AFL) Technical Whitepaper}}.
\newblock
  \bibinfo{howpublished}{\url{http://lcamtuf.coredump.cx/afl/technical_details.txt}}.
\newblock
\newblock
\shownote{{Accessed: 2019-09-06}.}


\end{thebibliography}

\appendix

\setcounter{table}{0}
\renewcommand{\thetable}{A\arabic{table}}

\clearpage

\section{Bugs and Reports}

\begin{table}[H]
\captionof{table}{The bugs injected into \sys, and the original bug reports. Of
the \sysbugcount bugs, 78 bugs (66\%) have a scope measure of one. Although most
single-scope bugs can be ported with an automatic technique, relying on such a
technique would produce fewer and lower-quality canaries. PoVs of $(\ast)$-marked
bugs are sourced from bug reports.}
\resizebox{0.48\linewidth}{!}{%
\rowcolors{2}{white}{rowgray}
\begin{tabular}[t]{llllcc}
\toprule
                          & Bug ID          & Report              & Class                             & PoV    & Scopes \\
\midrule
\multirow{\libpngbugn}{*}{\begin{sideways}libpng\end{sideways}}   & AAH001          & CVE-2018-13785      & Integer overflow, divide by zero  & \cmark & 1          \\
                          & AAH002$^\ast$          & CVE-2019-7317       & Use-after-free                    & \cmark & 4         \\
                          & AAH003          & CVE-2015-8472       & API inconsistency                 & \cmark & 2          \\
                          & AAH004          & CVE-2015-0973       & Integer overflow                  & \xmark & 1          \\
                          & AAH005$^\ast$          & CVE-2014-9495       & Integer overflow, Buffer overflow & \cmark & 1         \\
                          & AAH007          & (Unspecified)       & Memory leak                       & \cmark & 2         \\
                          & AAH008          & CVE-2013-6954       & 0-pointer dereference             & \cmark & 2         \\
\bottomrule
\end{tabular}
}
\hfill
\resizebox{0.48\linewidth}{!}{%
\rowcolors{2}{white}{rowgray}
\begin{tabular}[t]{llllcc}
\toprule
                          & \textbf{Bug ID} & \textbf{Report} & \textbf{Class}                & \textbf{PoV} & \textbf{Scopes} \\ \midrule
\multirow{\opensslbugn}{*}{\begin{sideways}openssl\end{sideways}} & AAH054          & CVE-2016-2842   & OOB write                    & \xmark & 5          \\
                          & AAH055          & CVE-2016-2108   & OOB read                     & \cmark & 5         \\
                          & AAH056          & CVE-2016-6309   & Use-after-free               & \cmark & 1         \\
                          & AAH057          & CVE-2016-2109   & Resource exhaustion (memory) & \xmark & 2          \\
                          & AAH058          & CVE-2016-2176   & Stack buffer overread        & \xmark & 2          \\
                          & AAH059          & CVE-2016-6304   & Resource exhaustion (memory) & \xmark & 3          \\
                          & MAE100          & CVE-2016-2105   & Integer overflow             & \xmark & 1          \\
\bottomrule
\end{tabular}
}
\label{tab:all-bugs}
\end{table}

\begin{table}
\centering

\caption{Mean bug survival times---both \textbf{R}eached
and \textbf{T}riggered---over a 24-hour period, in \textbf{s}econds,
\textbf{m}inutes, and \textbf{h}ours. Bugs are sorted by ``difficulty'' (mean
times). The best performing fuzzer is highlighted in green (ties are not
included).}
\label{table:eval}

\begin{adjustbox}{width=\linewidth}
\rowcolors{3}{rowgray}{white}
\begin{tabular}{l|rr|rr|rr|rr|rr|rr|rr|rr}
\toprule
                                    &
\multicolumn{2}{c|}{moptafl}        &
\multicolumn{2}{c|}{honggfuzz}      &
\multicolumn{2}{c|}{afl++}          &
\multicolumn{2}{c|}{afl}            &
\multicolumn{2}{c|}{aflfast}        &
\multicolumn{2}{c|}{fairfuzz}       &
\multicolumn{2}{c|}{symccafl}       &
\multicolumn{2}{c}{\colcombined} \\
Bug ID                              & R   & T
                                    & R   & T
                                    & R   & T
                                    & R   & T
                                    & R   & T
                                    & R   & T
                                    & R   & T
                                    & R   & T \\
\midrule
 AAH037 & 10.00s                       & 20.00s                       & 10.00s                       & {\cellcolor{hlgreen}} 10.00s & 10.00s                    & 45.50s                        & 5.00s                        & 15.00s                      & 5.00s                        & 15.00s                          & 5.00s                         & 15.00s                      & 10.00s                         & 25.50s                         & 7.86s            & 20.86s             \\
 AAH041 & 15.00s                       & 21.00s                       & 10.00s                       & {\cellcolor{hlgreen}} 10.00s & 15.00s                    & 48.00s                        & 10.00s                       & 15.00s                      & 10.00s                       & 15.00s                          & 10.00s                        & 15.00s                      & 15.00s                         & 30.00s                         & 12.14s           & 22.00s             \\
\bottomrule

\end{tabular}
\end{adjustbox}

\end{table}

\begin{table}\ContinuedFloat
\centering

\caption{Mean bug survival times (cont.). None of these bugs were triggered by
the \sysevalcount evaluated fuzzers.}

\begin{adjustbox}{width=\linewidth}
\rowcolors{3}{rowgray}{white}
\begin{tabular}{l|rr|rr|rr|rr|rr|rr|rr|rr}
\toprule
                                    &
\multicolumn{2}{c|}{moptafl}        &
\multicolumn{2}{c|}{honggfuzz}      &
\multicolumn{2}{c|}{afl++}          &
\multicolumn{2}{c|}{afl}            &
\multicolumn{2}{c|}{aflfast}        &
\multicolumn{2}{c|}{fairfuzz}       &
\multicolumn{2}{c|}{symccafl}       &
\multicolumn{2}{c}{\colcombined} \\
Bug ID                              & R   & T
                                    & R   & T
                                    & R   & T
                                    & R   & T
                                    & R   & T
                                    & R   & T
                                    & R   & T
                                    & R   & T \\
\midrule
 AAH054 & 10.00s                       & 24.00h                       & 5.00s                        & 24.00h                       & 10.00s                    & 24.00h                        & 5.00s                        & 24.00h                      & 5.00s                        & 24.00h                          & 5.00s                         & 24.00h                      & 10.00s                         & 24.00h                         & 7.14s            & 24.00h             \\
 MAE105 & 10.00s                       & 24.00h                       & 5.00s                        & 24.00h                       & 10.00s                    & 24.00h                        & 5.00s                        & 24.00h                      & 5.00s                        & 24.00h                          & 5.00s                         & 24.00h                      & 10.00s                         & 24.00h                         & 7.14s            & 24.00h             \\
\bottomrule

\end{tabular}
\end{adjustbox}

\end{table}

\begin{table}
\centering

\caption{Mean bug survival times over a 7-day period.}
\label{table:eval_7d}

\begin{adjustbox}{width=\linewidth}
\rowcolors{3}{rowgray}{white}

\begin{tabular}{l|rr|rr|rr|rr|rr|rr|rr|rr}
\toprule
                                    &
\multicolumn{2}{c|}{moptafl}        &
\multicolumn{2}{c|}{honggfuzz}      &
\multicolumn{2}{c|}{afl++}          &
\multicolumn{2}{c|}{afl}            &
\multicolumn{2}{c|}{aflfast}        &
\multicolumn{2}{c|}{fairfuzz}       &
\multicolumn{2}{c|}{symccafl}       &
\multicolumn{2}{c}{\colcombined} \\
Bug ID                              & R   & T
                                    & R   & T
                                    & R   & T
                                    & R   & T
                                    & R   & T
                                    & R   & T
                                    & R   & T
                                    & R   & T \\
\midrule
 AAH037 & 10.00s                        & 20.00s                        & 15.00s                       & {\cellcolor{hlgreen}} 15.00s  & 10.00s                       & 45.50s                        & 10.00s                      & 20.00s                       & 10.00s   & 20.00s                          & 10.00s                        & 21.00s                       & 10.00s                         & 25.50s                         & 10.71s           & 23.86s             \\
 AAH041 & 15.00s                        & 21.00s                        & 15.00s                       & {\cellcolor{hlgreen}} 15.00s  & 15.00s                       & 48.00s                        & 15.00s                      & 20.50s                       & 15.00s   & 20.00s                          & 15.00s                        & 21.00s                       & 15.00s                         & 30.00s                         & 15.00s           & 25.07s             \\
\bottomrule
\end{tabular}
\end{adjustbox}

\end{table}

\begin{table}\ContinuedFloat
\centering

\caption{Mean bug survival times (cont.).}

\begin{adjustbox}{width=\linewidth}
\rowcolors{3}{rowgray}{white}
\begin{tabular}{l|rr|rr|rr|rr|rr|rr|rr|rr}
\toprule
                                    &
\multicolumn{2}{c|}{moptafl}        &
\multicolumn{2}{c|}{honggfuzz}      &
\multicolumn{2}{c|}{afl++}          &
\multicolumn{2}{c|}{afl}            &
\multicolumn{2}{c|}{aflfast}        &
\multicolumn{2}{c|}{fairfuzz}       &
\multicolumn{2}{c|}{symccafl}       &
\multicolumn{2}{c}{\colcombined} \\
Bug ID                              & R   & T
                                    & R   & T
                                    & R   & T
                                    & R   & T
                                    & R   & T
                                    & R   & T
                                    & R   & T
                                    & R   & T \\
\midrule
 AAH054 & 10.00s                        & 168.00h                       & 10.00s                       & 168.00h                       & 10.00s                       & 168.00h                       & 10.00s                      & 168.00h                      & 10.00s   & 168.00h                         & 10.00s                        & 168.00h                      & 10.00s                         & 168.00h                        & 10.00s           & 168.00h            \\
 AAH011 & 10.00s                        & 168.00h                       & 10.00s                       & 168.00h                       & 10.00s                       & 168.00h                       & 10.00s                      & 168.00h                      & 10.00s   & 168.00h                         & 10.00s                        & 168.00h                      & 10.00s                         & 168.00h                        & 10.00s           & 168.00h            \\
\bottomrule
\end{tabular}
\end{adjustbox}

\end{table}

\end{document}